\documentclass[aps, reprint, twocolumn, floatfix,superscriptaddress,amsmath,amssymb,apsrev]{revtex4-1}
\usepackage{tabularx}
\usepackage{gensymb,wasysym, braket}
\usepackage{bm}
\usepackage{epsfig,subfigure}
\usepackage[normalem]{ulem}
\usepackage{multirow}
\usepackage{graphicx}
\usepackage{color}
\usepackage{amsfonts}
\usepackage{exscale}
\usepackage{amsbsy}
\usepackage{soul}

\graphicspath{{./Fig/}}
\usepackage[colorlinks,urlcolor=blue,linkcolor=blue,anchorcolor=blue,citecolor=blue]{hyperref}

\begin{document}
\begin{abstract}	
The antiferromagnetic spin-$1/2$ Heisenberg model on a kagome lattice is one of the most paradigmatic models in the context of spin liquids, yet the precise nature of its ground state is not understood. 
We use  large scale density matrix normalization group simulations (DMRG) on infinitely long cylinders  and find indications for the formation of a gapless Dirac spin liquid.
First, we use adiabatic flux insertion to demonstrate that the spin gap  is much smaller than estimated  from previous DMRG simulation.
Second, we find that the momentum dependent excitation spectrum, as extracted from the DMRG transfer matrix, exhibits Dirac cones that  match those of a $\pi$-flux free fermion model (the parton mean-field ansatz of a $U(1)$ Dirac spin liquid).  \end{abstract}
\title{Signatures of Dirac cones in a DMRG study of the Kagome Heisenberg model}
\author{Yin-Chen He}
\affiliation{Max-Planck-Institut f\"{u}r Physik komplexer Systeme, N\"{o}thnitzer Str. 38, 01187 Dresden, Germany}
\affiliation{Department of Physics, Harvard University, Cambridge, MA 02138, USA}
\affiliation{Kavli Institute for Theoretical Physics, University of California, Santa Barbara, CA 93106, USA}
\author{Michael P. Zaletel}
\affiliation{Station Q, Microsoft Research, Santa Barbara, CA 93106, USA}
\affiliation{Kavli Institute for Theoretical Physics, University of California, Santa Barbara, CA 93106, USA}
\author{Masaki Oshikawa}
\affiliation{Institute for Solid State Physics, University of Tokyo, Kashiwa 277-8581, Japan}
\affiliation{Kavli Institute for Theoretical Physics, University of California, Santa Barbara, CA 93106, USA}
\author{Frank Pollmann}
\affiliation{Max-Planck-Institut f\"{u}r Physik komplexer Systeme, N\"{o}thnitzer Str. 38, 01187 Dresden, Germany}
\affiliation{Kavli Institute for Theoretical Physics, University of California, Santa Barbara, CA 93106, USA}
\maketitle

\tableofcontents
\section{Introduction}
Understanding the ground state of the antiferromagnetic spin-$1/2$ Heisenberg model on a kagome lattice  (KAH)  has proved to be one of the most vexed issues in quantum magnetism~\cite{Sachdev1992,Chalker1992,Hastings2000,Singh2007,Ran2007,Hermele2008,Evenbly2010,Yan2011,Lauchli2011,Nakano2011,Depenbrock2012,Jiang2012,Messio2012,Iqbal2011,Iqbal2013,Iqbal2014,Iqbal2015,Nishimoto2013,He2015c,Laeuchli2015,Mei2016,Jiang2016,Liao2016}.
The KAH is one of the simplest models with strong frustration, and is a reasonable starting point for understanding various layered magnets such as Herbertsmithite \cite{Helton2007, Han2012}.
The possibility of a quantum spin liquid (QSL)~\cite{Anderson1973} on the KAH was proposed more than two decades ago~\cite{Sachdev1992},
and was more recently confirmed numerically through density matrix renormalization group (DMRG) simulations~\cite{Yan2011}.
However, there is a multitude of possible QSLs~\cite{SavaryBalentsReview, Zhou2016}, and despite tremendous efforts, the precise nature of the QSL for KAH remains unknown. Thus we will refer to it simply as the \emph{kagome spin liquid}.

Currently the most promising candidates are the gapped $Z_2$ spin liquid~\cite{ReadChakraborty89, Wen1991,Sachdev1992,Moessner2001} ( $Z_2$ SL) and the gapless $U(1)$ Dirac spin liquid (DSL) \cite{Hastings2000}. 
Recent DMRG studies appear to support a gapped $Z_2$ SL scenario \cite{Yan2011,Depenbrock2012, Jiang2012}.
However, characteristic properties of the gapped $Z_2$ SL, e.g. four-fold topological degeneracy on a torus and fractional statistics of spinons, have not been observed. 
On the other hand, there are several  indications favoring a DSL over a $Z_2$ SL. 
First, extensive variational Monte Carlo studies suggest a DSL \cite{Ran2007,Iqbal2011,Iqbal2013,Iqbal2014,Iqbal2015}.
Second, the kagome spin liquid was found to be proximate to a chiral spin liquid stabilized by the addition of longer ranged spin exchange interactions \cite{He2014,Gong2014}, with indications the transition between them is continuous \cite{He2015c}.
While there is no known theory that could describe a continuous transition between a $Z_2$ SL and the chiral spin liquid~\cite{Barkeshli2013,Zaletel2015}, such a transition occurs naturally if the kagome spin liquid is a DSL.
Third, a recent theoretical work~\cite{He2015a} suggests a DSL is a natural possibility by investigating a lattice gauge theory  formulation\cite{Nikolic2005,He2015} of the easy-axis (XXZ) kagome model \cite{He2015c,Laeuchli2015}. 

Most of the early experimental studies on candidate materials, such as Herbertsmithite, suggested a gapless spin liquid scenario based on spin susceptibility and specific heat measurements \cite{Helton2007}. 
However, the effective model for those materials is thought to be more complicated than the KAH, and it was difficult to determine if these gapless signatures were  intrinsic properties of the kagome spin liquid or were due to magnetic impurities.
A recent NMR study of Herbertsmithite found evidence for a finite spin gap ($0.03J \sim 0.07 J$) by using the Knight shift to filter out impurity contributions \cite{Fu2015}.

In this paper, we revisit the kagome spin liquid problem using the DMRG~\cite{White1992,White1993,McCulloch2008} method, which remains one of the most unbiased and powerful numerical methods to deal with this problem.%
We systematically investigate the energy gap and excitation spectrum of the kagome spin liquid phase using extensions to the previously used algorithms:
(i) We provide new insight into the heavily debated spin gap issue of the KAH by computing its dependence on boundary conditions, and show that the spin gap from our DMRG simulations is consistent with a gapless QSL (e.g. DSL).
(ii) We obtain the momentum-resolved spectrum of correlation lengths of the KAH, which is closely related to the excitation spectrum \cite{Zauner2014}.
In particular, this spectrum shows signatures of Dirac cones at the locations expected for a U(1) gapless DSL \cite{Hastings2000}.
The method we use here can also be directly applied to explore QSL phases in other lattice models.

The paper is organized as follows.
We begin by reviewing some promising spin liquid candidates and previous DMRG studies in Sec.~\ref{sec:review}.
We then discuss the expected behavior of various QSLs when placed on the cylinder geometry in Sec. \ref{sec:QSL_twist}.
We present our numerical DMRG data in Sec.~\ref{sec:KAH_results}.
First we show that the spin gap drops significantly compared to previously reported values as we twist the boundary conditions.
Then we extract the momentum dependent excitation spectrum from the DMRG transfer matrix. 
The triplet excitations reveal a Dirac cone structure, with the Dirac point located at the M point of the Brillouin zone, as expected for a $\pi$-flux DSL.
We finally conclude with a summary and discussion in Sec.~\ref{sec:con}.


\section{Brief Review: kagome spin liquids}\label{sec:review}

\subsection{Previous DMRG studies}
Yan et al. \cite{Yan2011} performed an extensive DMRG study of the kagome Heisenberg model on cylinders with circumference sizes from $L_y=2$ (e.g. YC4) to $L_y=6$ (e.g. YC12) unit cells.
Most importantly, it was found that the ground state is a symmetric spin liquid state that has a much lower energy than competing valence bond crystal states \cite{Singh2007,Evenbly2010}. 
By performing a careful finite size scaling analysis on different geometries, an energy per site of $ E_0 = -0.4379(3)$ was estimated.
Several observations were made regarding the nature of the spin liquid state:
(i) The ground state is a spin singlet protected by a small spin gap to the lowest lying spin-1 state. 
A spin gap  $\Delta_{S=1}=0.125(9)J$, with $J$ being the strength of the exchange interaction, was found for the $L_y=6$ (unit cells) cylinder (XC12-2).
This gap is smaller than the gap of  $\Delta_{S=1}=0.164J$ extracted from earlier exact diagonalization studies of a $36$ site cluster \cite{Waldtmann1998}.
(ii) The singlet gap, separating the ground-state from the lowest spin-0 excited state, is estimated to be much smaller, $\Delta_{S=0}=0.054(9)$ for the $L_y=6$ (unit cells) cylinder (XC12-2). 
This differs strongly from the exact diagonalization results that estimate the singlet gap to be less than $0.01J$ \cite{Waldtmann1998}.
The difference is attributed to strong finite size effects for the small clusters used for the exact diagonalization studies.
(iii) The ground state is found to have a very short correlation length.
The findings of Yan et al. were further corroborated by the SU(2)-invariant DMRG study of Depenbrock, et al. \cite{Depenbrock2012}.
It was further argued  based on the entanglement properties that the spin liquid is likely a $Z_2$ spin liquid \cite{Depenbrock2012, Jiang2012}. 

\subsection{Parton constructions for kagome spin liquids}
For the discussion that follows, it will prove very useful to have a picture of the competing phases within the language of the  fermionic parton construction.
Here, we briefly review the parton construction for a 2D plane \cite{Wen2002, SavaryBalentsReview,Zhou2016};  in Sec. \ref{sec:QSL_twist}, we discuss the novel phenomena which arise when a QSL is wrapped onto a cylinder.
	
In the parton construction, the physical spin operator is expressed as a bilinear of a fictitious $S=1/2$ fermion, $\hat{\mathbf{S}}_i = \sum_{\sigma, \sigma'} f^\dagger_{i, \sigma}  \mathbf{S}_{\sigma, \sigma'} f_{i, \sigma'}$.
In order to reproduce the correct properties for $\hat{\mathbf{S}}_i$, one enforces the constraint $1 = \sum_{\sigma}  f^\dagger_{i, \sigma}  f_{i, \sigma}$.
This introduces an extensive redundancy: for example, the spin operator is left invariant under U(1) `gauge transformations' $f_{i, \sigma} \to e^{i \phi_i} f_{i, \sigma}$ (in fact, there is actually a larger SU(2) redundancy).
In the resulting effective field theory, the fermionic partons are coupled to an emergent gauge field which both enforces the constraint and implements the microscopic redundancy as a gauge invariance. 

	Rather than elaborating on the field theory, we use the parton construction as a variational ansatz for the ground state and its low-lying excitations. Letting $\ket{\textrm{MF}}$ be an ansatz free-fermion wavefunction for the $f$, we can obtain a spin-1/2 wavefunction by projecting onto single occupancy,
\begin{align}
\ket{\Psi} = \prod_i n_i (2 - n_i) \ket{\textrm{MF}} =  \mathcal{P}_G \ket{\textrm{MF}}.
\end{align}
Various non-trivial spin-liquid phases result by choosing $\ket{\textrm{MF}}$ to be the ground state of certain free-fermion $H_{\textrm{f}}$, as will be discussed below.

The parton picture simultaneously suggests an ansatz for the excitations.
If $\gamma_\sigma$ is a low-lying excitation of $H_{\textrm{f}}$, with $\sigma = \uparrow / \downarrow$, we can obtain a corresponding ansatz for a low-lying excitation of the spin-system,
\begin{align}
\ket{\gamma_\sigma} =   \mathcal{P}_G \gamma_\sigma^\dagger \ket{\textrm{MF}}.
\end{align}
This excitation is called a fermionic `spinon' excitation, since it carries a S=1/2 moment.
Consequently the spinon is a topological excitation, e.g., it cannot be made by acting with the local $\mathbf{S}_i$ operators in some patch.
Triplet excitations are obtained from the two-spinon states.
If the mean-field description is well behaved, the energy of the ansatz excitation $\ket{\gamma_\sigma}$ should have some qualitative relation to the energy of $\gamma$ in $H_\textrm{f}$; for example, the location of band minima or gapless points.

We now turn to the free-fermion ansatz $H_{\textrm{f}}$ thought to be relevant for the kagome spin liquid.
\subsubsection{Gapped $\mathbb{Z}_2$ spin-liquid}
	We obtain a gapped $Z_2$ spin-liquid \cite{ReadChakraborty89, Sachdev1992,Wen1991} by supposing that the partons form a BCS superconductor: $H_\textrm{f} = - \sum_{\langle i, j\rangle , \sigma } t_{i, j} f_{i \sigma}^\dagger f_{j \sigma}    + \Delta_{ij} f_{i \uparrow} f_{i \downarrow} + h.c.  - \mu N $.
Various choices of sign-structures for the $t, \Delta$ (corresponding, for example, to patterns of $\pi$-flux through plaquettes) in fact lead to eight known gapped Z$_2$-SLs, which differ in how the crystal symmetries act on the spinon excitations \cite{Sachdev1992, Wang2006, Lu2011, Qi2016, Lu2016}.
In all eight cases the BdG particles $\gamma_\sigma$ are gapped, leading to a spin gap. 
There is also a second type of excitation in a $\mathbb{Z}_2$ spin-liquid: a $\pi$-flux of the emergent gauge field which is coupled to the partons (the other gauge excitations are at very high energy due to the Higgs mechanism).
An ansatz for this $\pi$-flux (alias `vison') excitation is given by appropriately modifying the pairing $t_{ij} \to - t_{ij}, \Delta_{ij} \to - \Delta_{ij}$ along a semi-infinite line to mimic $\pi$-flux piercing the plane at the endpoint,  finding the resulting free-fermion ground state, and Gutzwiller projecting.
The $\pi$-flux is also a topological excitation, but it does not carry spin.
The existence of the $\pi$-flux  leads to important phenomena on the  cylinder we will return to.

\subsubsection{$U(1)$-Dirac spin-liquid}
In the U(1)-Dirac spin-liquid, we choose a hopping-only ansatz \cite{Hastings2000, Ran2007, Iqbal2011, Iqbal2013, Iqbal2014,Iqbal2015}: 
\begin{equation}	
H_\textrm{f} = - \sum_{\langle i, j\rangle , \sigma } t_{i, j} f_{i \sigma}^\dagger f_{j \sigma} + h.c.  - \mu N.\label{eq:mean}
\end{equation}
The uniform nearest-neighbor ansatz $t_{ij} = 1$ is a poor choice for the kagome model, since it leads to a flat valence band: with an odd number of sites in the unit cell at a density of $n = 1$ per site, this band would be half full.
Instead, one considers the hopping pattern shown in Fig.~\ref{fig:geometry}(b), which has $\pi$-flux piercing the hexagons.
This $\pi$-flux ansatz results in a halved magnetic Brillouin zone with two Dirac cones at momenta $Q = (\pi/2, \pi/2), Q' = (-\pi/2, -\pi/2)$.
Combined with spin, there are $N_f = 4$ gapless Dirac cones, and hence a vanishing spin gap.
Since the emergent gauge field is not `Higgsed', there are also low energy gauge fluctuations, and the resulting effective theory is essentially $N_f = 4$ QED$_3$.
	
\subsubsection{Chiral spin-liquid}
	The chiral spin-liquid is also a hopping-only ansatz \cite{Hastings2000}, but we choose the $t_{ij}$ to have phases such that the occupied bands have a total Chern number of $C=1$ per spin species. 
From the point of view of the DSL, it is obtained by turning on the same-sign mass term, $\bar{\psi}_{Q, \sigma} \psi_{Q, \sigma}$, on all $N_f = 4$ Dirac cones.
By choosing $C=1$, the ansatz breaks both time-reversal $T$ and reflection $P$, though their combination $PT$ is preserved. 
Generically this leads to a non-vanishing expectation value for the chiral order parameter $\mathbf{S}_i \cdot \mathbf{S}_j \times \mathbf{S}_k$, where $i, j, k$ are three nearby sites.
Since the Chern band is gapped, there is a finite energy cost for spinon excitations, and the emergent gauge field is gapped by the effective Chern-Simons term.
The Chern-Simons term leads to a spin-Hall coefficient of $\sigma^{(s)}_{H} = \tfrac{1}{2}$.

Note that the Dirac-SL can be considered the `parent state' for many of these spin-liquids: the Z$_2$ gapped states are obtained by turning on some superconducting terms, while the CSL arises if a  staggered $B$-field spontaneously forms in addition to the $\pi$-flux.
This picture gives a natural scenario for continuous phase transitions into these neighboring phases.

\section{Spin liquids on a cylinder \label{sec:QSL_twist}}
Before providing numerical DMRG results for the kagome spin liquid, it is worthwhile to pause and consider what one should expect for a spin liquid in the DMRG simulation, which compactifies the kagome lattice into a cylinder.
Since spin liquids are topological phases, they display a number subtle features on a cylinder.

\subsection{Kagome cylinder geometries}

\begin{figure}
\includegraphics[width=0.49\textwidth]{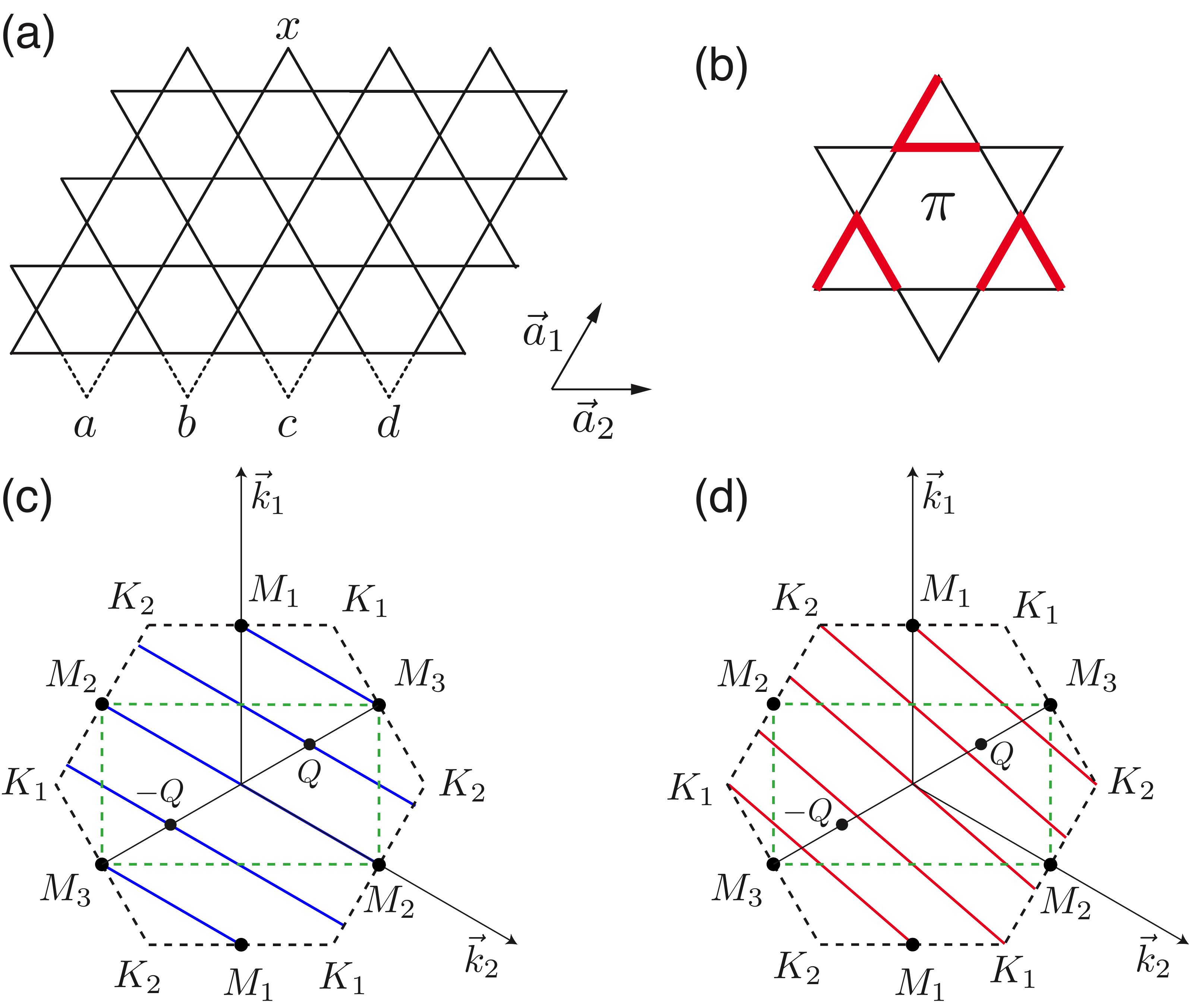} \caption{\label{fig:geometry} (a) Illustration of the kagome lattice with the width of $L_y=4$ unit cells. The different geometries, YC8-0, YC8-2, YC8-4, YC8-6, correspond to identifying the site $x$ with the site $a$, $b$, $c$ or $d$ respectively. (b) The $\pi$-flux state, where each hexagon has $\pi$-flux. We chose a specific gauge, in which the red bond has $s_{ij}=-1$, and the black bond has $s_{ij}=1$. 
The different geometries (here (c) YC8-0 and (d) YC8-2) actually correspond to different ways of cutting the Brillouin zone. The $M$ points are (labeled by $(k_1, k_2)$), $M_1=(\pi, 0)$, $M_2=(0, \pi)$ and $M_3=(\pi, \pi)$. The two Dirac points of the DSL are at $\pm Q$. The DSL has a halved magnetic BZ (dashed), since it is a $\pi$-flux ansatz.
 }
\end{figure}
A cylinder is defined by identifying sites that differ by some $\vec r$, which defines the different compactifications.
We work with the YC2n-2m cylinders,  where $\vec x$ is identified with $\vec x + n \vec a_1 - m \vec a_2$ and $\vec a_1, \vec a_2$ are the kagome Bravais vectors defined in Fig.~\ref{fig:geometry}(a).
We will have $n > m$, so we call $L_y=n$ the `circumference' of the cylinder, while $m$ amounts to a shift of the most naive identification $m=0$.
We note that YC2n-0, YC2n-n are actually the YC2n and XC2n geometries introduced in Ref. \cite{Yan2011}.

Since one direction of the system is compactified, the momentum along that direction is  discretized, specifically
\begin{equation} \label{eq:discrete}
n  k_1 +m  k_2= 0, \quad \mod 2\pi.
\end{equation}
Note that the other direction is infinite, so $k_1$ and $k_2$ can take continuous values so long as they satisfy the above relation.
The different geometries (different $n$ and $m$) thus provide different cuts through the Brillouin zone, as shown in Fig. \ref{fig:geometry}(c)-(d).

A second parameter one can tune is the boundary condition $\theta$ of the spins around the circumference, the same way one would measure spin-stiffness. 
We obtain a twist boundary condition  by modifying Heisenberg coupling according to $S_{x, y}^+ S_{x', y+1}^- \rightarrow e^{i\theta / L_y} S_{x, y}^+ S_{x', y+1}^-$. By analogy to flux threading,  we call  $\theta$ the spin-flux in the cylinder , as shown in Fig. \ref{fig:SL_twist} (a).
\subsection{Gapped SL on a cylinder}

Gapped spin liquids have fractionalized quasiparticles that have non-trivial statistics (e.g.,  the fermionic spinon). 
A direct consequence of such fractionalization is the topological degeneracy on a torus; there is one ground state per quasiparticle type \cite{ReadChakraborty89, Wen1990b}.
Precisely the same degeneracy arises for a long (infinite) cylinder -- one can  picture the ends of the cylinder as identified into a torus at infinity. The energy splitting between these `topological sectors' is exponentially small in the circumference.

In the Z$_2$-SL, the topological degeneracy can be  understood in terms of the boundary conditions of the fermionic partons.
While the boundary conditions of the spins are set by $\theta$, the fermionic spinons are coupled to an emergent gauge field, and the flux `$\phi$' of this gauge field through the cylinder effectively changes their boundary conditions.
In a superconductor, $\pi$-flux is invisible to the condensate, so periodic (PBC) and anti-periodic (APBC) will give nearly degenerate energies - the splitting should decay exponentially in the cylinder circumference vs. coherence length.

\begin{figure}
\includegraphics[width=0.49\textwidth]{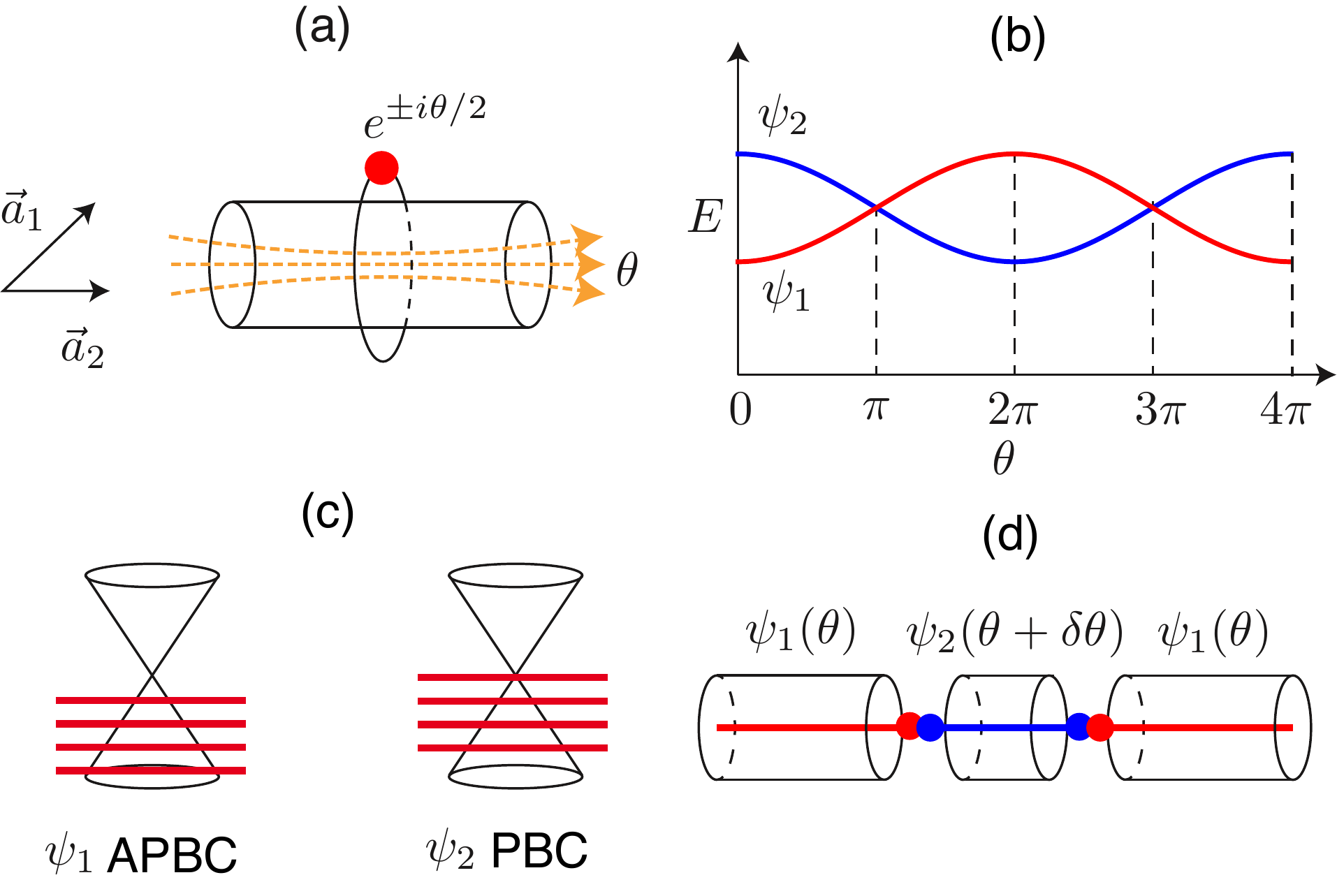}
\caption{\label{fig:SL_twist} The cartoon picture for a spin liquid under the twist. (a) Spinon feels half twist boundary/flux due to the fractionalization. (b) The response of the energy under the twist.  (c) Two topological sectors of a DSL, with one gapped sector and one gapless sector. (d) The collapse of the twist simulation will have quasiparticle emerging as the domain wall.}
\end{figure}

There is an interesting interplay between the topological degeneracy and spin-flux $\theta$.
Spin rotations $S^{\pm}_i \to e^{\pm i \Omega }S^{\pm}_i$ are implemented on the partons as $f_{i, \sigma} \to e^{i \Omega \sigma/2} f_{i, \sigma}$, where $\sigma=\pm 1$ corresponds to spin-up and spin-down components (the eigenvalue of $2S^z$). As a consequence, the spin flux $\theta$ will change the boundary conditions of the up/down partons by $\pm \theta/2$.
A superconductor has no spin-stiffness, so the change in the energy will again be exponentially small in the circumference of the cylinder. However, at $\theta = 2 \pi$,  $H_\textrm{f}$ (in Eq.\eqref{eq:mean}) does not return to itself: because they carry half-integer spin, the boundary conditions for the $f_{\sigma}$ have been changed by $\pm 2 \pi / 2 \equiv \pi$. 
Equivalently, threading $2 \pi$-spin flux adiabatically exchanges between the PBC and APBC topological sectors, as shown in Fig. \ref{fig:SL_twist}(b).
Note that the location of the crossing is pinned to $\theta = \pi$ by time-reversal, where it acts by exchanging the two topological sectors.

In summary, there is a striking signature of  a gapped $\mathbb{Z}_2$ spin liquid: as spin-flux is inserted, there should be an exponentially small change ($L_x\mathcal{O}[e^{-L_\textrm{cyl} / \xi}]$) in the ground state energy, and at $\theta = 2\pi$ the ground state should \emph{not} return to itself, but instead should thread the emergent $\pi$-flux that corresponds to the other topological degenerate groundstate \cite{Laughlin1981,Oshikawa2000b}.

\subsection{Dirac spin liquid on a cylinder}
\label{sec:DiracCylinder}
Like the $\mathbb{Z}_2$-SL, the DSL has an emergent gauge field which can dynamically change the boundary conditions of spinons (including either PBC or APBC).
However, because of the semi-metallic parton band structure the response is quite different: any twist of the parton boundary conditions will shift the energy by $v_F L_x  / L_y^2$, where $v_F$ is the spinon velocity, due to the energy of the filled parton bands below the Dirac point.
In particular, a boundary condition in which the allowed $k$-modes avoid the Dirac point will be the lowest in energy, since this effectively opens up a gap of order $v_F / L_y$.
Thus when a Dirac spin liquid is placed on a narrow cylinder, the gauge field will generically adjust to open up a gap. Therefore, a numerical observation of a non-vanishing gap on a single long cylinder generally does not rule out the possibility of a gapless DSL.

Here, flux threading can be used to find fingerprints of a DSL on a cylinder.
As before, adiabatically threading spin-flux $\theta$ through the cylinder  twists the up/down parton boundary conditions by $\pm \theta/2$ as shown in  Fig.~\ref{fig:SL_twist}(a).
This twist is in addition to the (shared) emergent flux $\phi$.  For symmetry reasons, the internal gauge flux will either be $\phi = 0, \pi$.
As we thread spin flux, the energy spectrum will again generically look like Fig.~\ref{fig:SL_twist}(b), but in contrast to a gapped SL the splitting is only \emph{algebraically} small.
Beyond the crossing, it will become difficult to adiabatically track the state due to the large splitting and small gap (since the allowed $k$-modes becomes close to the Dirac point); at some  point $\pi$-emergent flux will enter to partially cancel the spin-flux, at which point there will be a discontinuous jump in the ground state energy of the DMRG simulation \cite{He2014a}.

The expected behavior of the gap depends on the geometry, which we consider in more detail at the mean-field level in order to interpret our numerical results.
At spin-flux $\theta$, the momenta of the partons $f_{\uparrow / \downarrow}$ on a YC2n-2m cylinder are restricted to
\begin{equation}
n k_1+m k_2= \phi \pm \theta/2 \quad \mbox{mod } 2\pi 
\end{equation}
where $\phi=0, \pi$ is the emergent gauge flux.
The $\pm$ correspond to spinons $f_{\uparrow}, f_{\downarrow}$ respectively. There are two classes of cylinders which behave very differently with $\theta$:
\begin{itemize}
\item{Type I cylinder: YC2n-4k.}
For YC2n-4k cylinders, when $\theta = 0$  the flux $\phi = \pi$ avoids both Dirac points, while both are present for $\phi = 0$. In this case, we would (naively) expect the gap to decrease like $ (2 \pi - \theta) / L_x$  until adiabaticity is lost after $\theta > \pi$ and the emergent $\pi$-flux tunnels in.
Hence we can't force the system to go gapless.
\item{Type II cylinder: YC2n-(4k+2).}
In contrast,  the  YC2n-(4k+2) cylinder becomes gapless at $\theta = \pi$. Here, for $\phi = 0$ the $\uparrow Q$ and $\downarrow Q'$ components are gapless, while for $\phi = \pi$ the $\downarrow Q$ and $\uparrow Q'$ are.
All other values of spin-flux have a gap which should decrease as  $ |\pi - \theta| / L_x$.
Thus at $\theta = \pi$  two of the four Dirac fermions are present \emph{regardless} of the emergent flux, and hence we can force the system to go gapless.
\end{itemize}

\section{Numerical results}
\label{sec:KAH_results}
In the following, we will discuss the numerical results for the KAH obtained using DMRG simulations on infinite cylinders (iDMRG)~\cite{McCulloch2008}. 
Besides the  nearest-neighbor Heisenberg interactions, we  also include a small-second neighbor interaction for some simulations,
\begin{equation}
H=J_1 \sum_{\langle i j \rangle} \vec S_i \cdot \vec S_j + J_2 \sum_{\langle \langle i j \rangle\rangle}  \vec S_i \cdot \vec S_j.
\end{equation}
We will study the behavior of the spin gap and transfer matrix spectrum (an analog of excitation spectrum) as we adiabatically twist the boundary conditions (spin-flux $\theta$).
We implement the adiabatic twist by using the previous DMRG wavefunction as the initial step for the next $\theta$ value  \cite{He2014a}.
The ground state energy during the insertion is provided in the Appendix.
Note that while DMRG generally finds the absolute ground state, when passing adiabatically through a level crossing there may be a small regime in which the DMRG follows the higher, metastable level until `tunneling' into the lower state, a phenomena we encounter below.
We find that the behavior of all the geometries can be grouped into the two types discussed in Sec.~\ref{sec:DiracCylinder}.
For the Type I cylinders (YC8-0 and YC8-4), the  adiabaticity fails at $\theta\approx 4\pi/3$, and the simulation collapses to the other topological sector with lower energy; \cite{He2014a} note for $ \pi < \theta \lesssim 4\pi/3$  we are tracking the higher, metastable state, not the absolute ground state.
For Type II (YC6-2, YC8-2, YC10-2, YC12-2, YC8-6), the adiabaticity fails around $\theta\approx \pi$, at which point we find an instability of the kagome spin liquid towards an ordered state.
For a DSL, this instability corresponds to a spontaneous generation of mass $i\bar \Psi \sigma^3 \mu^3 \Psi$ (see the Appendix Sec.~\ref{sec:instability} for more details).

\subsection{Spin gap}
\begin{figure}
\includegraphics[width=0.49\textwidth]{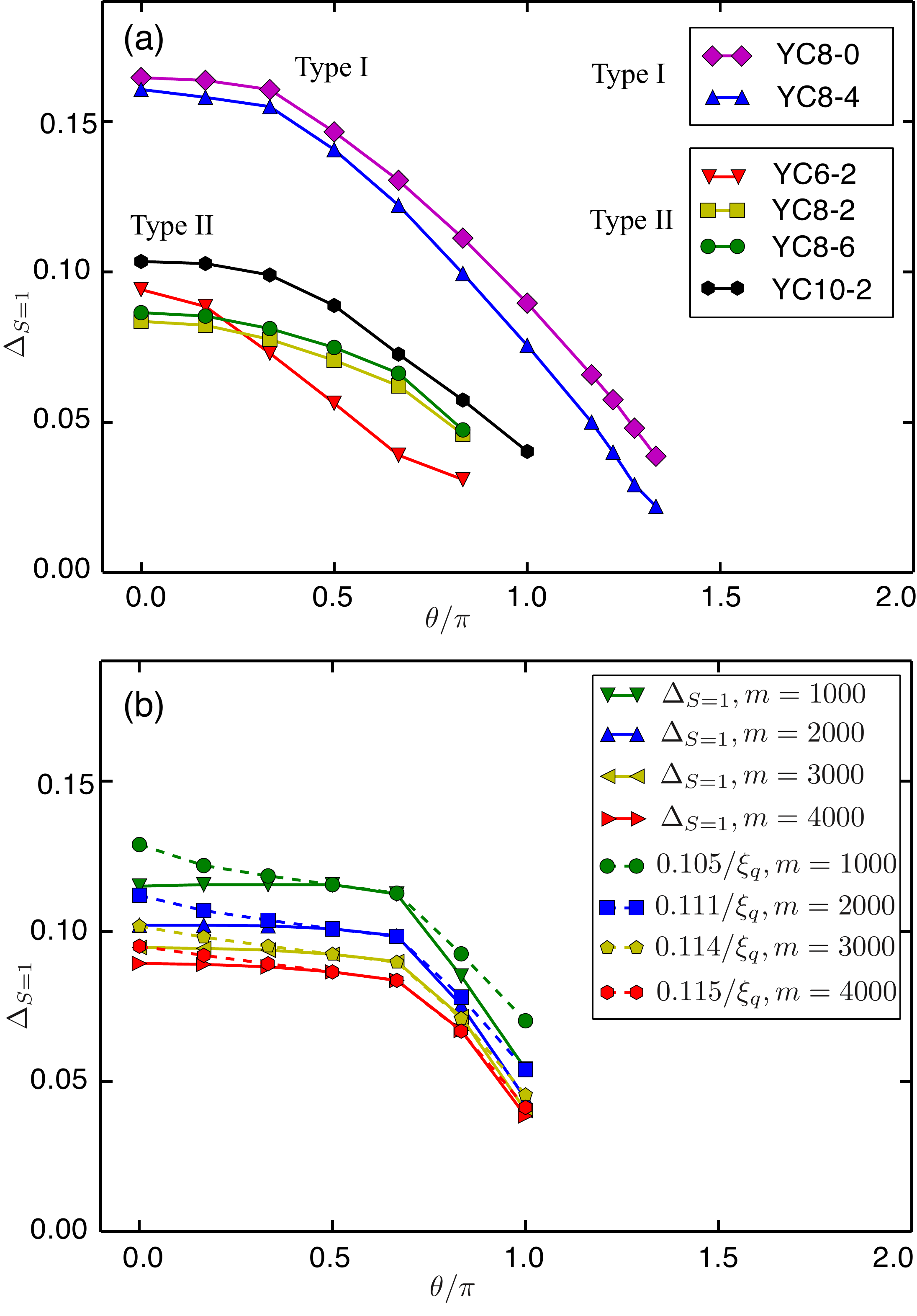} \caption{\label{fig:gap} 
(a) Upper bound on the spin gap $\Delta_{S=1}$ under the twist boundary condition $\theta$ for $J_2=0$; the estimate is obtained using DMRG bond dimension  $m=4000$ and using the lowest-energy topological sector.
Data ends at the failure of adiabaticity.
The qualitative behavior depends on the type of  cylinder (I or II), which in a DSL would be gapless at $\theta = 2 \pi, \pi$ respectively.
(b) Dependence of the spin gap $\Delta_{S=1}$ (solid line) and $S^z=1$ correlation length $\xi$ (dashed line) on the spin flux $\theta$  and DMRG bond dimension $m$. Data is taken with $J_2=0.05$ on the Type II  YC8-2 cylinder. 
Generally the estimated gap decreases with the bond dimension $m$;  the larger the system size is, the more likely we will overestimate on the spin gap.}
\end{figure} 
The spin (triplet) gap $\Delta_{S=1}$ is obtained by creating a $S^z=1$ excitation in the bulk of the cylinder and then calculating the energy difference to the $S^z=0$ sector \cite{Zaletel2013, He2014} (see  Appendix Sec. \ref{sec:algorithm} for details).
We stress here that the spin gap is different from the (extensive) energy splitting between the different topological sectors.

Fig.~\ref{fig:gap}(a) provides raw data for the spin gap as function of the twist angle $\theta$ for DMRG bond dimension $m=4000$.
Generally the spin gap decreases with increasing bond dimension, as seen in Fig.~\ref{fig:gap}b, and thus the data shown provides only an upper bound for the spin gap.
When $\theta=0$,  the spin gap is rather large, for example YC8-0 and YC8-4 have a spin gap of  $\Delta_{S=1}\approx 0.15J$, completely consistent with previous DMRG simulations \cite{Yan2011,Depenbrock2012}.
Strikingly, the spin gap shows a significant decrease when $\theta$ is increased, and right before the failure of adiabaticity, the spin gap drops to a very small value $\Delta_{S=1}\approx 0.02J - 0.04J$.
The approximately linear decrease of the gap for larger twist angles $\theta$ is  suggestive of a DSL gapless spin liquid.
As discussed in the previous section, the spin gap for a gapped spin liquid should have an exponentially small ($\mathcal{O}(e^{-L_y / \xi}$) dependence on the twist angle $\theta$, though of course $\xi$ can be large.

We emphasize that, based on our data alone, we cannot conclude whether the gap will vanish in the thermodynamic limit.
For large enough $L_y$, the spin gap of a DSL at $\theta=0$ should decrease with the circumference size $L_y$ as $\Delta_{S=1} \sim v_F/L_y$.
This behavior has not been observed in previous studies \cite{Yan2011, Depenbrock2012}, nor in our current simulations.
On the one hand, this could be an artifact of  the finite bond dimension $m$ of DMRG simulations;  finite $m$ tends to overestimate the spin gap (Fig. \ref{fig:gap}(b)), necessitating a careful extrapolation in $1 / m^\alpha$, where $\alpha$ is an unknown power.
Since $m$ must increase exponentially with circumference to achieve the same accuracy, it becomes very challenging to accurately extract the spin gap for large circumferences.
Alternatively, for  small circumferences it is unclear whether the gap will follow the naive mean field expectation $\Delta_{S=1} \sim v_F/L_y$; a possible reason is discussed in Sec.~\ref{sec:exc_interp}.

\subsection{Transfer matrix spectrum \label{sec:spectrum}}
We examine the transfer matrix spectrum on the infinite cylinder (i.e., the correlation-length spectrum) and compare the KAH with a free fermion $\pi$-flux model.
In the ansatz wavefunction of the DMRG (matrix product states), the correlation functions of charge-$q$ operators (e.g., the spin-spin correlations $C_{S=1}(r)=\langle S^+_0S^-_r\rangle_{\textrm{con}}$) can be expanded as a sum of exponentials,
\begin{equation}
C_q(r) = \sum^{m_q}_{j=1} \alpha_{q,j} \lambda_{q,j}^{r}.
\end{equation}
Here $r$ is the distance along the long direction of the cylinder, $q$ is a quantum number, $m_q$ is a bond dimension, and $\lambda_{q,j}$ are eigenvalues of the DMRG `transfer matrix' \cite{Schollwoeck2011}. 
Using the quantum numbers $q$, we can distinguish, for example, a triplet excitation from a singlet excitation (or more precisely, a $S^z=1$ excitation from a $S^z=0$ excitation).
The eigenvalues $\lambda_{q,j}= e^{ ik_{q,j}-\xi^{-1}_{q,j}}$ have a real part,  corresponding to a correlation length  $\xi_{q,j}$, and an imaginary part, corresponding to a momentum  $k_{q,j}$.    
The largest, $\xi_q = \textrm{max}(\xi_{q,j})$, bounds the correlation length of all charge-$q$ operators.

\begin{figure}
\includegraphics[width=0.49\textwidth]{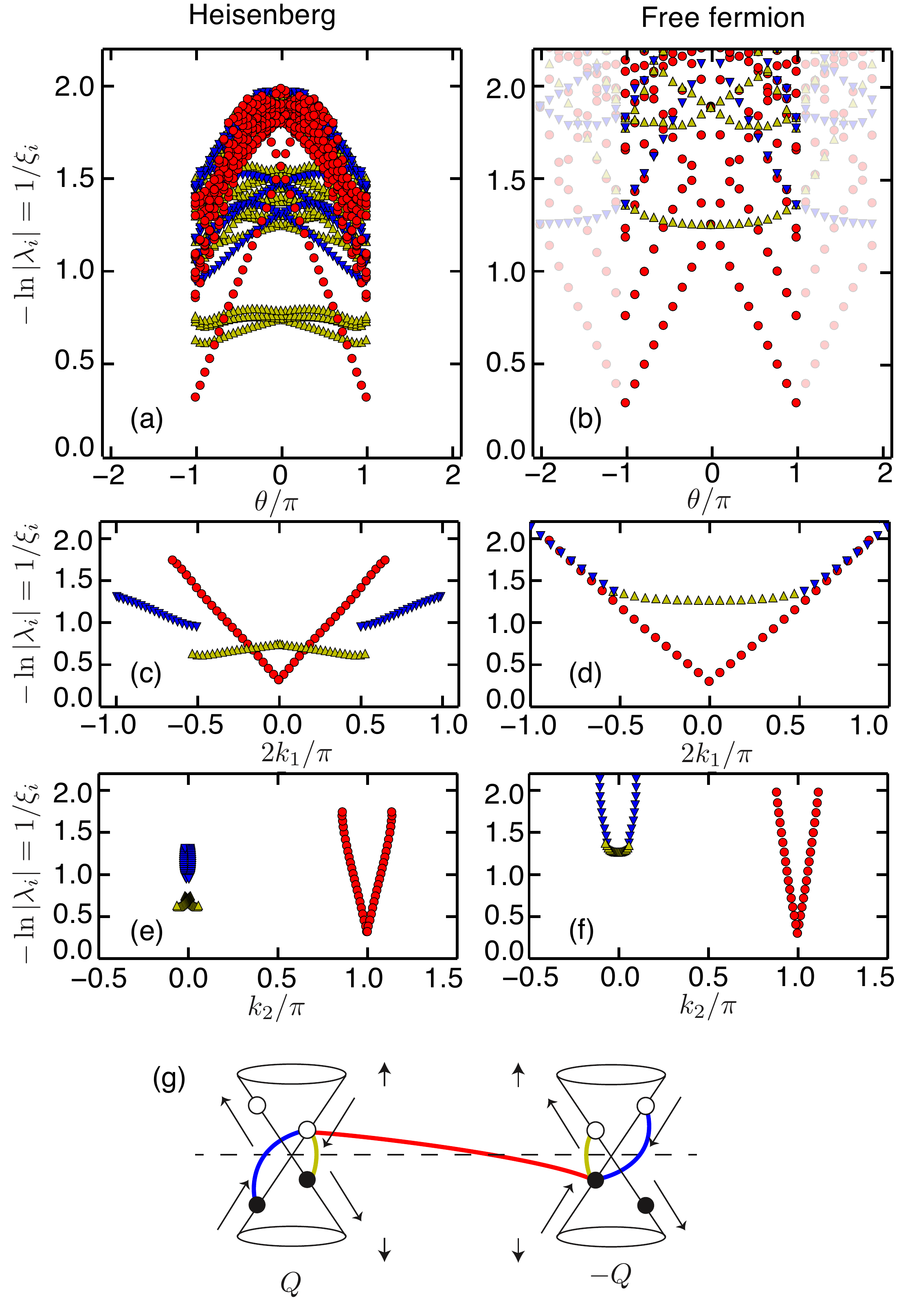} \caption{\label{fig:S1_twist} The transfer matrix spectrum in the $S^z=1$ sector (roughly the triplet channel) for the kagome Heisenberg model (a), (c), (e) and free fermion model (b), (d), (f).
The vertical axis is the inverse of correlation length $1/\xi$, which can be considered as the gap $\Delta$ of the excitations, up to a pre-factor $\Delta=v_s /\xi$.
The horizontal axis denotes (a), (b) the twist angle $\theta$; (c), (d) momentum $2k_1$; (e), (f) momentum $k_2$.
The cylinder we show here is the YC8-2 cylinder for both case, and the truncation error of DMRG is around $2\times 10^{-6}$, which corresponds to bond dimension $m=6000$ for the kagome Heisenberg model and $m=250$ for the free fermion model.
For the kagome Heisenberg model, we also include a small $J_2=0.05$.
(g) Three different types of particle-hole excitations. The arrows represent the direction of the movement of the discretized momentum under the twist boundary conditions.
}
\end{figure}
A recent work by Zauner et al. \cite{Zauner2014} pointed out a relation between the  energy spectrum of the physical excitations  and the spectrum of the transfer matrix.
A more familiar statement is that the largest correlation lengths $\xi_q$ set an upper bound for the lowest excitation gaps $\Delta_q$ (up to a factor), and for a Lorentz invariant systems it holds that $\Delta_q \propto 1/\xi_q$.
The corresponding $k_q$ gives the momentum of the excitation along the length of the cylinder. 
These relations actually hold nicely for the KAH at different twist angles, as demonstrated in Fig.~\ref{fig:gap}b.

Figures~\ref{fig:S1_twist}a-b show the $S^z=1$ transfer matrix spectrum of the KAH and the $\pi$-flux free fermion model Eq.~(\ref{eq:mean}) as function of the twist angle $\theta$.
We consider the Type II YC8-2 geometry, and included a small $J_2$ to stabilize the adiabatic twist up to $\theta=\pi$  (the Appendix shows results for other geometries).
The three different colors label three `bands:' we observe that the momenta $k_{S=1, j}$ cluster into three distinct groups, and we plot the largest several $\xi_{S=1, j}$  from each momenta group.
The momentum can be resolved into its lattice components $k_1, k_2$, providing an alternative way to plot the data shown in Figures~\ref{fig:S1_twist}c-f.

\subsubsection{Interpretation as DSL \label{sec:exc_interp}}
The KAH and free fermion spectra are remarkably similar. The excitation spectrum can be  understood based on the free fermion $\pi$-flux model.
$S^z=1$ excitations arise from particle-hole excitations near the Dirac points; a momentum $p-q$ spin flip takes the form $S^+(p-q)=f_{\uparrow}^\dag(p) f_\downarrow (q)$.
The $\pi$-flux state  has two Dirac points at $Q=(\pi/2, \pi/2)$ and $-Q=(-\pi/2, -\pi/2)$.
We group the particle-hole excitations into intra-valley forward (blue); intra-valley backward (yellow); and inter-valley forward (red), as illustrated in the cartoon \ref{fig:S1_twist}(g) (inter-valley backward scattering is  higher in energy).
The Dirac points are avoided on the $\theta=0$ YC8-2 cylinder, as shown in Fig. \ref{fig:geometry}(d), but as $\theta$ increases, the allowed momenta shift and eventually pass through the Dirac point; the $f_\uparrow$ and $f_\downarrow$ feel opposite flux, hence move oppositely.
As can be seen in the cartoon, this shift affects the three modes in a qualitatively different fashion.
The dispersion of the red mode follows a Dirac behavior and  becomes gapless at the twist angle $\theta=\pm \pi$.
In terms of momenta, the gapless point occurs at $(2k_1, k_2)=(0, \pi)$ as expected from the displacement between $Q$ and $Q'$.
The yellow mode has a constant energy under the twist angle $\theta$. The blue mode has similar response as the red mode, but remains gapped when the system hits the Dirac points ($\theta=\pi$).

The spectrum of the KAH and the $\pi$-flux free fermion model show surprisingly good agreement: (i) the red mode has a linear sharp Dirac cone structure; (ii)  the yellow mode is almost flat; (iii) the modes occur with the predicted momenta.
The qualitative difference between two models is that the yellow and blue modes in the KAH are lower  compared with the free fermion model.
It may that even though the DSL theory should have an emergent SU(4) symmetry in 2D, in the quasi-1D geometry  intra-valley interactions are stronger.

The existence of the renormalized flat yellow band also explains the `kink' in the $\theta$-dependence of the triplet gap $\Delta_{S=1}$: for small $\theta$ it drops below the linear red band, which then cross.
This implies that gaps obtained in previous DMRG studies, which all worked at $\theta = 0$, were probing the yellow intra-valley excitation.
Since the yellow band is subject to strong interaction effects, this may relate to the non-observation of $v_F / L_y$ gap scaling on accessible cylinders.

We want to remark that within our DMRG simulations the correlation spectrum of the KAH still has a finite ``gap" even at the Dirac point.
This is a necessary consequence of DMRG, since the finite bond dimension $m$ induces a finite correlation length.
We find the $\xi$ increase with $m$, as expected for a DMRG simulation of a critical system.
In fact, a similar behavior is found also in the free fermion model.
The correlation length estimated from DMRG (with $m=250$ in Fig. \ref{fig:S1_twist}) is finite even for the gapless free fermion model with an infinite correlation length. For a very large bond dimension ($m=3000$, see Appendix Fig. \ref{fig:free_app}), the correlation length becomes very large ($\xi \sim 1000$ sites),  supporting that the finite correlation length at the Dirac point (in Fig. \ref{fig:S1_twist}) is purely an artifact of small bond dimension.

\subsection{$S^z=0$ spectrum and scalar chirality}
\label{sec:chirality}
\begin{figure} 
\includegraphics[width=0.49\textwidth]{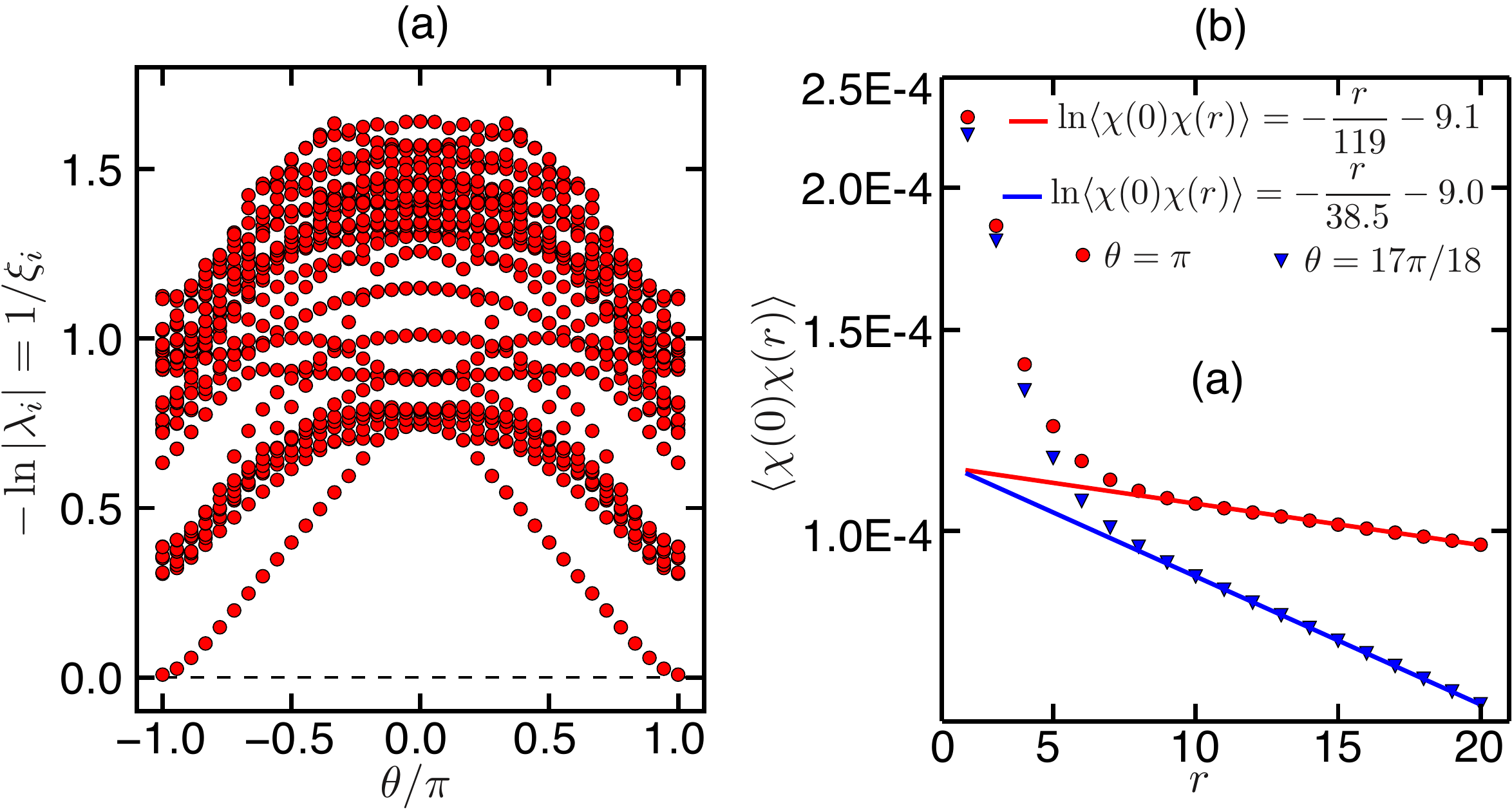} \caption{\label{fig:S0_twist} (a) The transfer-matrix spectrum of the kagome Heisenberg model in the $S^z=0$ sector versus the twist angle $\theta$.
Here we have YC8-2 cylinder, $J_2=0.05$, bond dimension $m=6000$.
(b) Correlation function of the scalar chirality $\langle \chi(0)\chi(r) \rangle$ at $\theta=\pi$ and $17\pi/18$. The correlation lengths from fitting $\langle \chi(0)\chi(r) \rangle$  are $\xi=119$ ($\theta=\pi$) and $\xi=38.5$ ($\theta=17\pi/18$), which are virtually identical with the correlation lengths obtained from the transfer matrix $\xi=121$ and $\xi=39$. }
\end{figure}

We now turn to the $S^z=0$ transfer matrix spectrum shown in Fig.~\ref{fig:S0_twist}(a), which  includes singlet excitations for which the interactions have a more drastic effect. 
Note that because our numerics do not explicitly preserve SO(3) symmetry,  the $S^z = 0$ spectrum contains both SO(3) singlets and elements of SO(3) multiplets. 
The correlation length in the $S^z=0$ sector shows a critical feature: at the the twist angle $\theta=\pi$, the $S^z=0$ correlation length $\xi \sim 100$ unit cells, which (holding fixed the bond dimension $m = 6000$) is much larger than the correlation length $\xi \sim 5$ in the $S^z=1$ sector.
This  effect is beyond a mean-field analysis, since for a free fermion, there will be no difference between $S^z=0$ and $S^z=1$ sectors.

The large correlation length in the $S^z=0$ sector arises from long-range correlations of the scalar chirality $\chi=\vec S_i \cdot (\vec S_j \times \vec S_k)$. 
To confirm this, in
Fig.~\ref{fig:S0_twist}(b), we show a fit for the correlation function of the scalar chirality, $\langle \chi(0) \chi(r)\rangle\sim e^{-r/\xi}$.
Indeed, the correlation length $\xi$ is almost identical to the largest correlation length obtained from the transfer matrix in the $S^z=0$ sector.
The large correlation length of the scalar chirality is somewhat surprising given our knowledge of $N_f = 4$ QED$_3$ ($U(1)$ DSL). 
The scalar chirality corresponds to the SU(4) singlet $\bar \psi \psi$, which was suggested to have higher scaling dimension than the SU(4) adjoint fermion bilinears \cite{Hermele2005erratum, Hermele2007}. Naively, an operator with a lower scaling dimension in a critical theory would give a larger correlation length in DMRG calculation with a finite bond dimension, in an apparent contradiction with our result.
However, our result does not necessarily indicate the scalar chirality has the lowest scaling dimension in the critical theory, since the magnitude of the correlation length is not directly equivalent to the scaling dimension of an operator.
Another possible explanation is that the cylinder geometry drastically changes the scaling analysis,  since (if it was stable) QED$_3$ dimensionally reduces to $N_f - 1$ coupled Tomonaga-Luttinger liquids~\cite{Sheng2009}, for which the scaling analysis may differ substantially from the 2D limit.
In addition, we note that the SU(4)-invariant mass $\bar \psi \psi$ generates the CSL ( it gives the bands a net Chern-number), and we know that the KAH is proximate to the CSL \cite{He2014,Bauer2014,Gong2014,He2015c}.
It was also suggested that the scalar chirality contains a monopole operator, which might have a  lower scaling dimension than the fermion bilinears \cite{Hermele2008}.

\section{Conclusion and discussion}
\label{sec:con}

We used large scale DMRG simulations to study the quantum spin liquid phase in the $S=1/2$ kagome antiferromagnetic Heisenberg model.
DMRG studies of  the KAH are most natural on a cylinder.
We point out that, even if a gapless QSL such as the Dirac spin liquid  is realized in the two-dimensional bulk limit, the system generally acquires a non-vanishing gap on the cylinder.
The predicted size of this gap depends sensitively on the geometry and boundary conditions of the cylinder,  complicating the finite size scaling analysis.
Thus the observation of a gap in the previous DMRG studies of the KAH, which is also reproduced in our study, might not rule out a  gapless QSL until the gap can be accurately measured for a sequence of `equivalent' geometries (e.g., YC8, YC12, YC16 $\cdots$), which is extremely challenging due to the exponential blow-up in DMRG bond dimension.

To better identify the nature of the QSL in the KAH using DMRG, we insert spin-flux through the cylinder, 
changing the boundary condition around the circumference. 
Using adiabatic flux insertion we find that the spin gap on the cylinder geometry is much smaller than estimated from previous DMRG simulations.
Second, we found that the momentum dependent excitation spectrum as estimated from the DMRG transfer matrix exhibits Dirac cones that agree well with the ones found for a $\pi$-flux free fermion model (the parton mean-field ansatz of a $U(1)$ Dirac spin liquid).
These findings suggest that the ground state of the KAH is a gapless DSL, instead of a gapped QSL such as the $\mathbb{Z}_2$ topological phase.
This is more in line with several recent
numerical~\cite{Ran2007,Nakano2011,Iqbal2011,Iqbal2013,Iqbal2014,Iqbal2015} and
analytical~\cite{He2015a} results obtained by methods other than DMRG.

However, we also have to make some cautious remarks.
While we found indications of gapless features in the system sizes that we can access, we cannot draw a definite conclusion for the thermodynamic limit.
Diverging correlation lengths  require extremely large bond dimensions for proper convergence of the gap, making it impossible to rule out the existence of a small but finite gap with a large correlation length.
In addition, fixing $\theta = 0$ the spin gap should eventually decrease as $v_F / L_y$, which has not yet been observed.
Within a DSL scenario, this could be a finite-size effect due to observed  strong renormalization of the intra-valley backward scattering, or it could be a numerical artifact of finite-bond dimension.
Finally, the present analysis does not reveal the nature of the gauge field, which could be either $U(1)$ or $\mathbb{Z}_2$.
Our result therefore is certainly not the final answer to the long-standing question on the KAH.
Nevertheless, our results strongly suggest that the ground state of the KAH is a DSL, which has many theoretical and experimental implications.

\section*{Acknowledgement}
We thank M. Barkeshli, Y.M Lu, M. Metlitski, A. Vishwanath and C. Wang for useful discussions.
MZ thanks D. Huse and S. White for collaboration on related work.
YCH thanks for the hospitality of Institute for the Solid State Physics, University of Tokyo, where this project was initiated.
YCH is supported by a postdoctoral fellowship from the Gordon and Betty Moore Foundation, under the EPiQS initiative, GBMF4306, at Harvard University.
This work was supported in part by JSPS Strategic International Networks Program No. R2604 ``TopoNet'' (YCH and MO),
JSPS Grant-in-Aid for Scientific Research (KAKENHI) No. 16K05469 (MO), and DFG via SFB 1143.
This work of MZ was performed in part at the Aspen Center for Physics, which is supported by National Science Foundation grant PHY-1066293.
YCH, MZ, MO and FP thank the hospitality of the Kavli Institute for Theoretical Physics, which is supported in part by the National Science Foundation under Grant No. NSF PHY-1125915

\appendix
\section{Numerical algorithm \label{sec:algorithm}}
\subsection{Transfer matrix and its spectrum}

\begin{figure}
\includegraphics[width=0.49\textwidth]{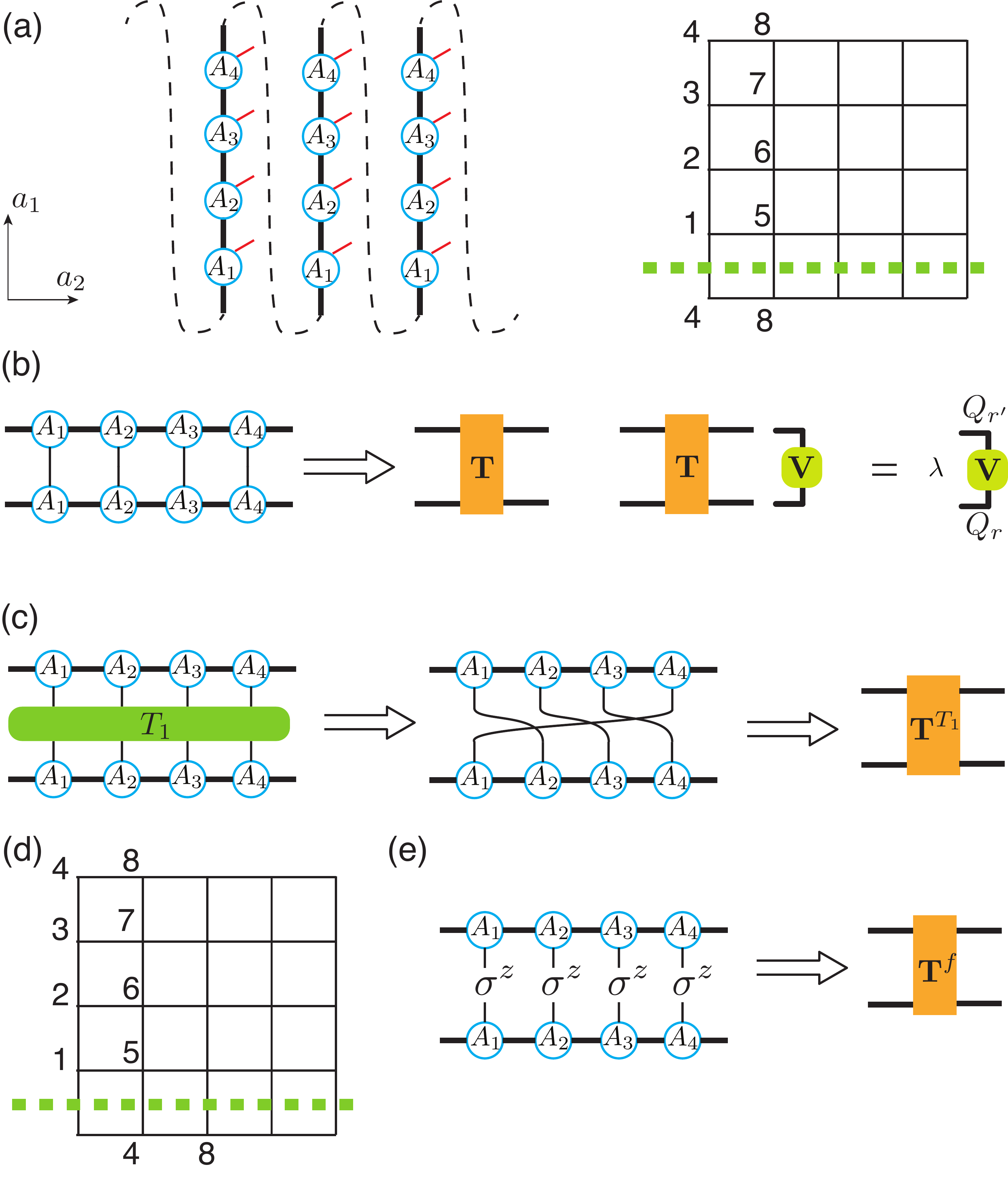}\caption{\label{fig:transfer_matrix}The graphical representation of (a) the MPS (b) the transfer matrix. The quantum number of the eigenvalue is determined by $q=Q_r-Q_{r'}$.
(c) The mixed transfer matrix to calculate the momentum $k_1$ of each Schmidt basis. (d) The -2 geometry for the square lattice, the periodic condition is taken by identifying the sites labeled by the same number. (e) The mixed transfer matrix for the fermionic excitation. }
\end{figure}

Numerically, we wrap a 2D lattice on a cylinder with one direction compactified with a small number of sites, and the other direction infinite.
We use the MPS to cover the cylinder in the fashion of a snake chain, as graphically shown in Fig.~\ref{fig:transfer_matrix}(a).
Along the compactified direction ($a_1$), the snake covering does not maintain the translational symmetry explicitly, thus one needs distinct MPS for each site.
On the other hand, the MPS is translation invariant and repeating along the direction $a_2$.
Then one can use the MPS (of the smallest repeating unit cell) to define the transfer matrix (TM), as shown in Fig. \ref{fig:transfer_matrix}(b).
With the TM, one can further calculate its eigenvalues $\lambda_{q,j}= e^{ ik_{q,j}-\xi^{-1}_{q,j}}$,  which has a real part, corresponding to a correlation length  $\xi_{q,j}$, and an imaginary part, corresponding to a momentum  $k_{q,j}$. 
$q$ is the quantum numbers, from which we can distinguish for example triplet excitation from singlet excitation.

Due to the snake covering, the MPS does not have translational invariance along the compactified direction ($a_1$).
However, the Hamiltonian still has translational invariance along $a_1$, hence the momentum $k_1$ along that direction can still be extracted.
The way to calculate $k_1$ is similar as calculating a global quantum number $q$, for which one needs to obtain the momentum $k_1$ of each Schmidt basis.  
Technically, we implement the translational operation $T_1$ along the direction $a_1$, and then obtains a mixed TM--$\mathbf T^{T_1}$, as shown in Fig. \ref{fig:transfer_matrix}(c).
The dominant eigenvector of $\mathbf T^{T_1}$ will be $\mathbf V_{\alpha, \beta}=\delta_{\alpha, \beta} e^{i k_\alpha}$, and $k_\alpha$ gives the momentum $k_1$ of each Schmidt basis.

We want to remark that for certain special geometries, for instance the YC2n-2, the momentum along the compactified direction cannot be extracted using the method discussed above.
The reason is because for those geometries, the momentum $k_1$ and $k_2$ are intertwined together.
A consequence is that one cannot define the mixed TM for the translation $T_1$.
For example we can consider the -2 geometry for the square lattice Fig. \ref{fig:transfer_matrix}(d). 
One can see that under the $T_1$ translation, the sites on one column doesn't go back to itself, instead some of the sites will go to  the neighboring column.
This is different from the normal geometry (see the left panel of Fig. \ref{fig:transfer_matrix}(a)), for which $T_1$ translation maps the sites on one column back to itself, hence the mixed TM--$\mathbf T^{T_1}$ can be well defined.
A benefit of the -2 geometry, however, is that the snake-fashion MPS is actually translational invariant under 2 sites, no matter how large the circumference of the cylinder is. 
Then one can actually use a MPS with 2-site structure (for the kagome lattice, it is 6 sites) to do the iDMRG simulation. 
With the momentum $k$ from the TM's eigenvalue, one can then obtain the momentum 
\begin{equation}
2k_1=k+2\theta/L_y, \quad\quad k_2=k L_y/2
\end{equation}
Here $L_y$ is the width of the cylinder, $\theta$ is the twist boundary condition.
Therefore, for the -2 geometry, one can still get $k_1$ and $k_2$, but $k_1$ has a $\pi$ ambiguity.

Before closing this section, it is worth to make a few remarks on fermionic systems. 
Usually to simulate a fermionic system, we first do a Jordan-Wigner transformation to obtain the corresponding bosonic model, and then simulate the bosonic model directly.
Therefore, to calculate the spectrum of the fermionic (e.g. single particle)  excitations, one should consider the mixed TM--$\mathbf T^f$, with a Jordan-Wigner string inserted, Fig. \ref{fig:transfer_matrix}(e).

\subsection{Algorithm of calculating the spin gap}
We use an algorithm that combines iDMRG and finite DMRG to calculate the spin gap. Firstly, we use iDMRG to obtain a converged wave-function of an infinite cylinder, then we cut the infinite cylinder into two halves, insert a $3\times L_y \times L_y$ sites into the system. The left ($L$) and right ($R$) semi-infinite cylinder can be considered as environment (boundary conditions), and we calculate the energy of ground state  $E_0(S_z=0)$, spin-1 sector $E_0(S_z=1)$ within the inserted $3\times L_y \times L_y$ cylinder. Finally, we obtain the spin gap $\Delta_{S=1}=E_0(S_z=1)-E_0(S_z=0)$. 

\begin{figure}
\centering
\includegraphics[width=0.3\textwidth]{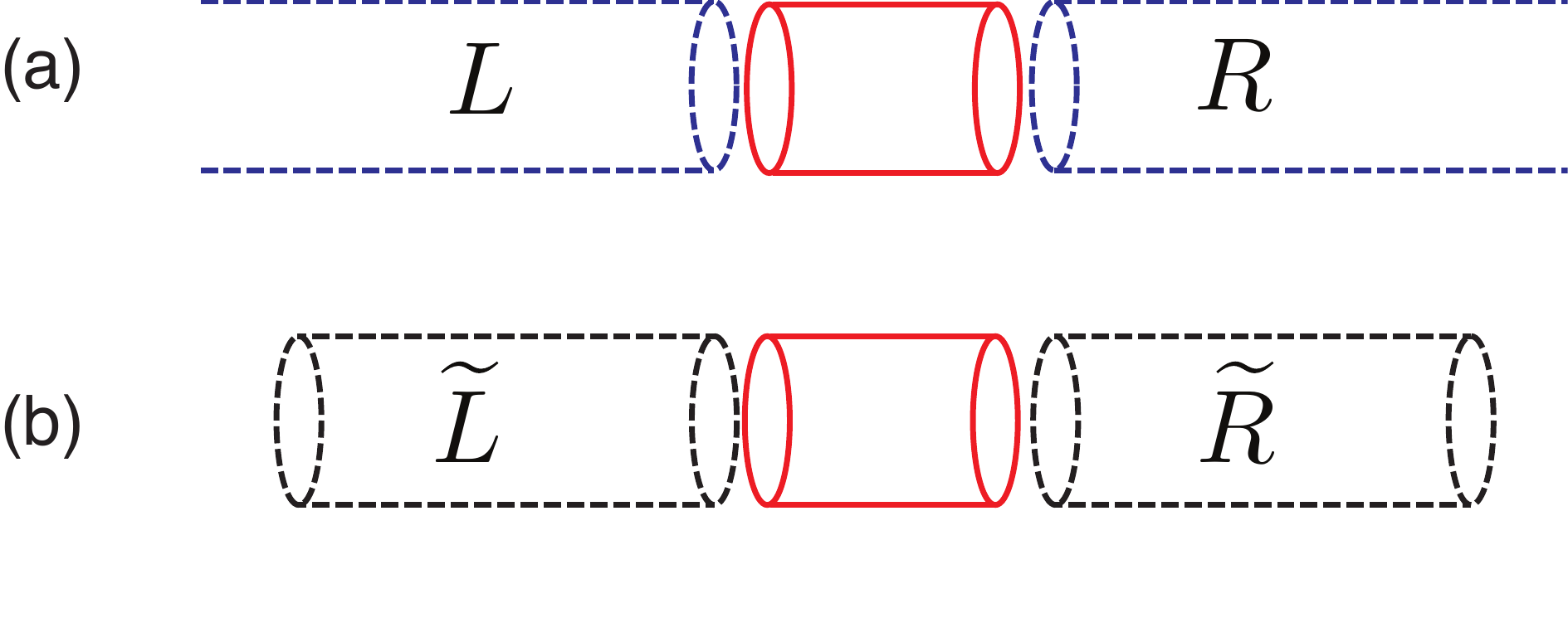} \caption{\label{fig:gap_algorithm} (a) iDMRG--finite DMRG combined algorithm (b) Finite DMRG algorithm.}
\end{figure}

This algorithm is similar as the one used in finite DMRG, where one obtains the spin (triplet) and singlet gap by sweeping in the middle of a finite cylinder to minimize the boundary effect. The only difference is that the boundary environment we use comes from the iDMRG simulation, while finite DMRG uses environment from finite cylinder simulation.

\section{Additional numerical data}
\subsection{Benchmark for free fermions\label{sec:free}}

\begin{figure}
\includegraphics[width=0.49\textwidth]{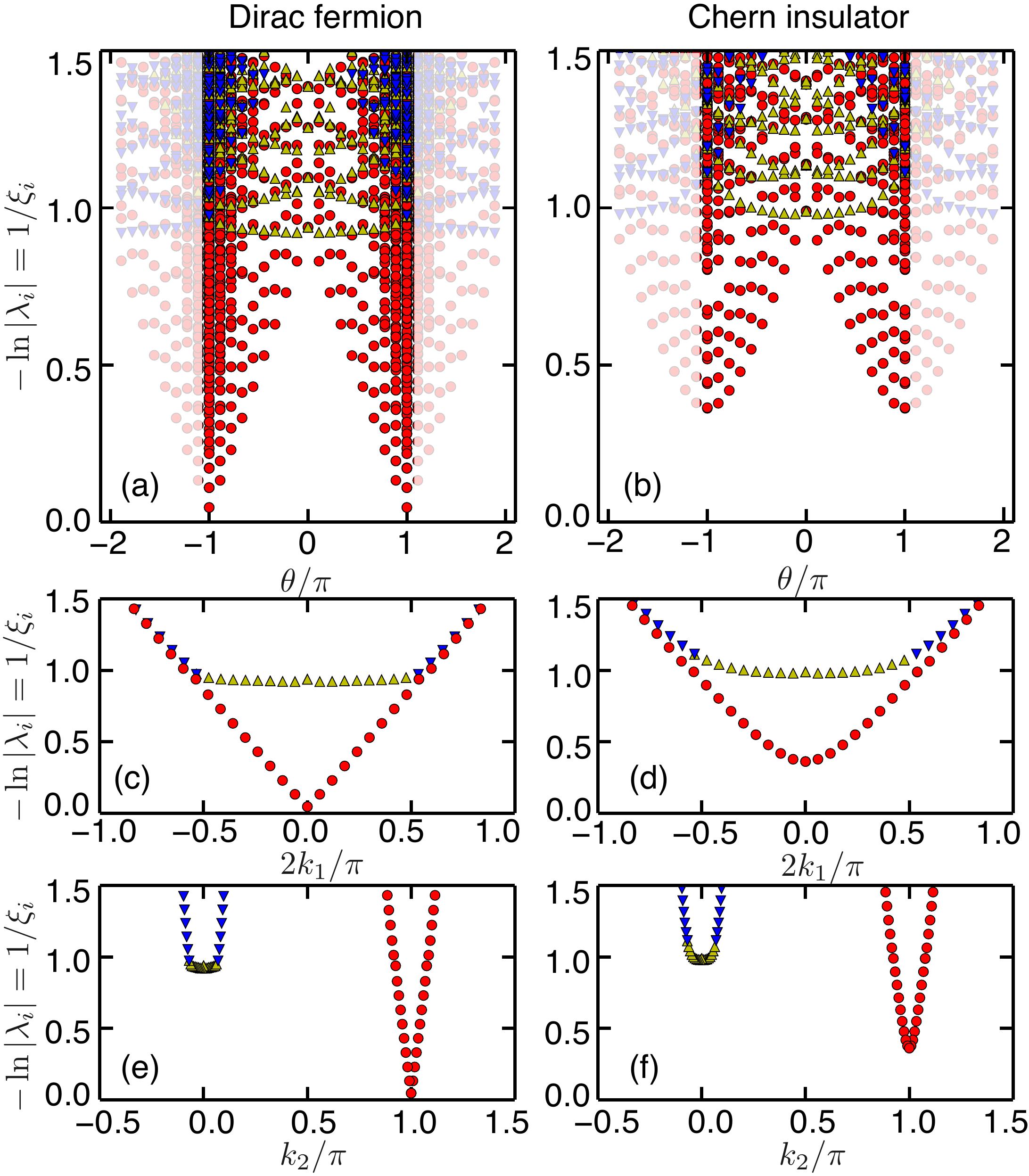} \caption{\label{fig:free_app} Transfer matrix spectrum for free  fermions (Eq. \ref{eq:mean}) on the YC8-2 cylinder with bond dimension $m=3000$. Here we compare the free Dirac fermion model and a Chern insulator: the inverse of correlation lengths $1/\xi$ versus  (a),(b) twist angle, (c), (d) momentum $2k_1$ and (e), (f) momentum $k_2$.
The Chern insulator we show here is obtained by assigning $\pi/25$ flux in the up and down triangles (of kagome), and $23\pi/25$ flux in the hexagon.}
\end{figure}

We begin with the transfer matrix spectrum of the free fermion model, the $\pi$-flux state on the kagome lattice Eq. \eqref{eq:mean}. 
Here we show the data with a larger bond dimension $m=3000$ (Fig. \ref{fig:free_app}), meanwhile we also compare the case that the groundstate is a Chern insulator.  It is clear that compared with the small bond dimension data ($m=250$ in the Fig. \ref{fig:S1_twist}(b),(d),(f)), the spectrum of free Dirac fermions (Fig.\ref{fig:free_app}(a), (c), (e)) is almost critical at the Dirac point.
This supports that the finite correlation length of the kagome Heisenberg model at the Dirac point, is simply an artifact of finite bond dimension in the DMRG simulations.
On the other hand, we also calculate the case of a Chern insulator, which clearly shows a parabola shape in the spectrum (Fig.\ref{fig:free_app}(b), (d), (f)).

\subsection{The kagome Heisenberg model}

\subsubsection{Spin-stiffness}

\begin{figure}
\includegraphics[width=0.49\textwidth]{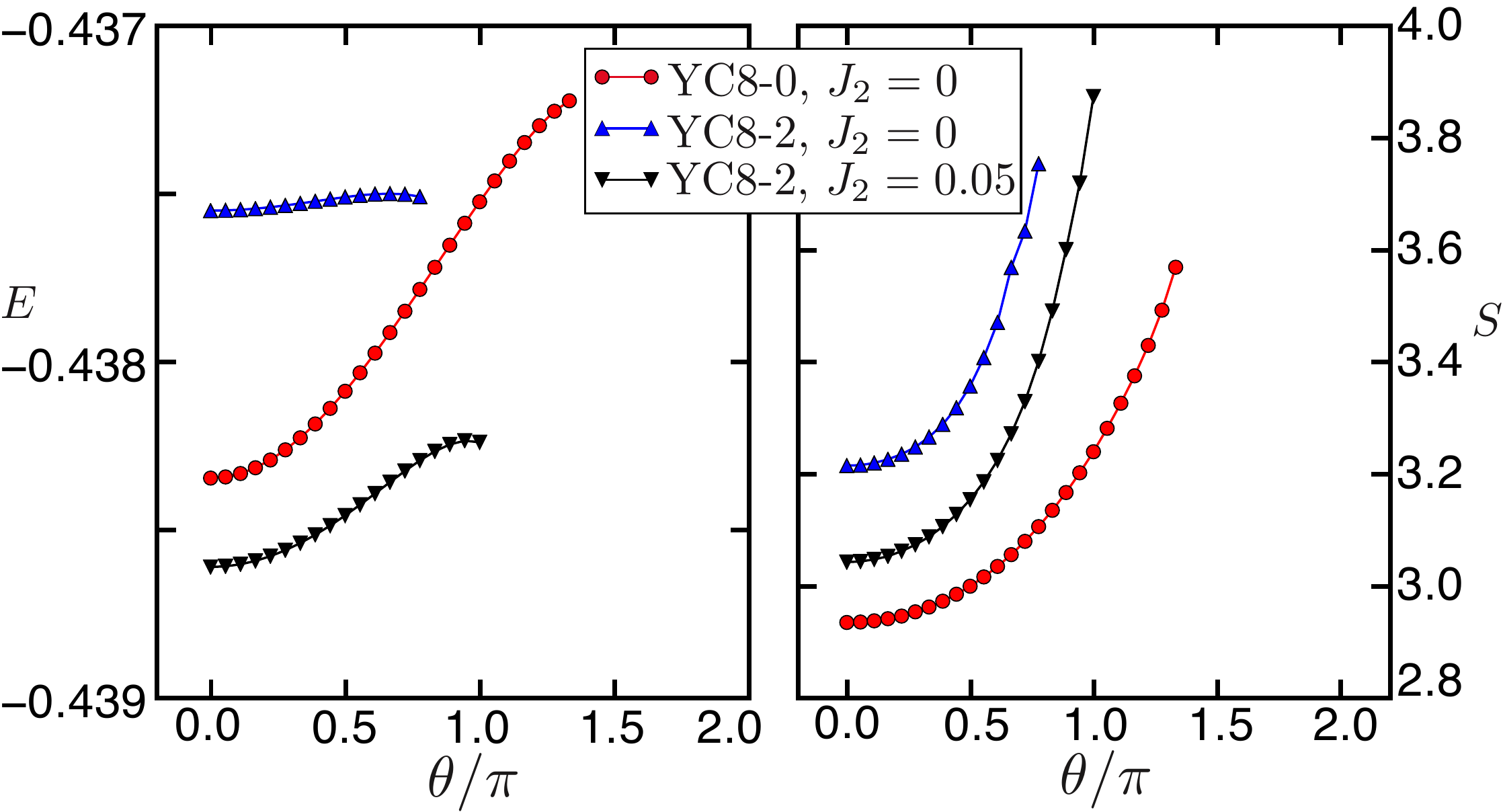}
\caption{\label{fig:kagome_twist} The energy and entanglement entropy of the KSL under flux insertion. Here the bond dimension is $m=6000$.}
\end{figure}

Fig. \ref{fig:kagome_twist} shows  the response of the ground state energy and entanglement entropy under the twist boundary conditions before the failure of adiabaticity.
Practically,  the adiabaticity can be checked by looking at the wave-function overlap($\approx 0.99$) between each two adjacent twist angles.
During the (adiabatic) twist process, the system remains in a spin liquid phase that preserves all the lattice symmetries.
Both the energy and entanglement entropy increase under the twist;  the increase in entanglement entropy is very significant, and may be underestimated by finite $m$. 
This behavior is consistent with the DSL.
It is worth noting that, for YC8-0, the twist trajectory is not symmetric about $\theta=\pi$, which is a clear signature of fractionalization; the state at $\theta=\pi$ is not time-reversal-invariant as well (also true for YC8-2).

\subsubsection{Transfer matrix}
We provide more data for the transfer matrix spectrum of the KSL.
We will show that by changing the system sizes and parameter point ($J_2$ interaction), the feature of a Dirac cone structure always exists. 
For example, the $S^z=1$ excitation shows a clear Dirac cone structure.
As before, the $S^z=0$ excitation is always much lower than the $S^z=1$ excitation.

As we described in the main text, the behavior of the system can be grouped into two types, YC2n-(4k+2) cylinder and YC2n-4k cylinder.
In the following, we will look at the two different classes separately.
\subsubsection{YC2n-(4k+2) cylinder}
\begin{figure}
\includegraphics[width=0.49\textwidth]{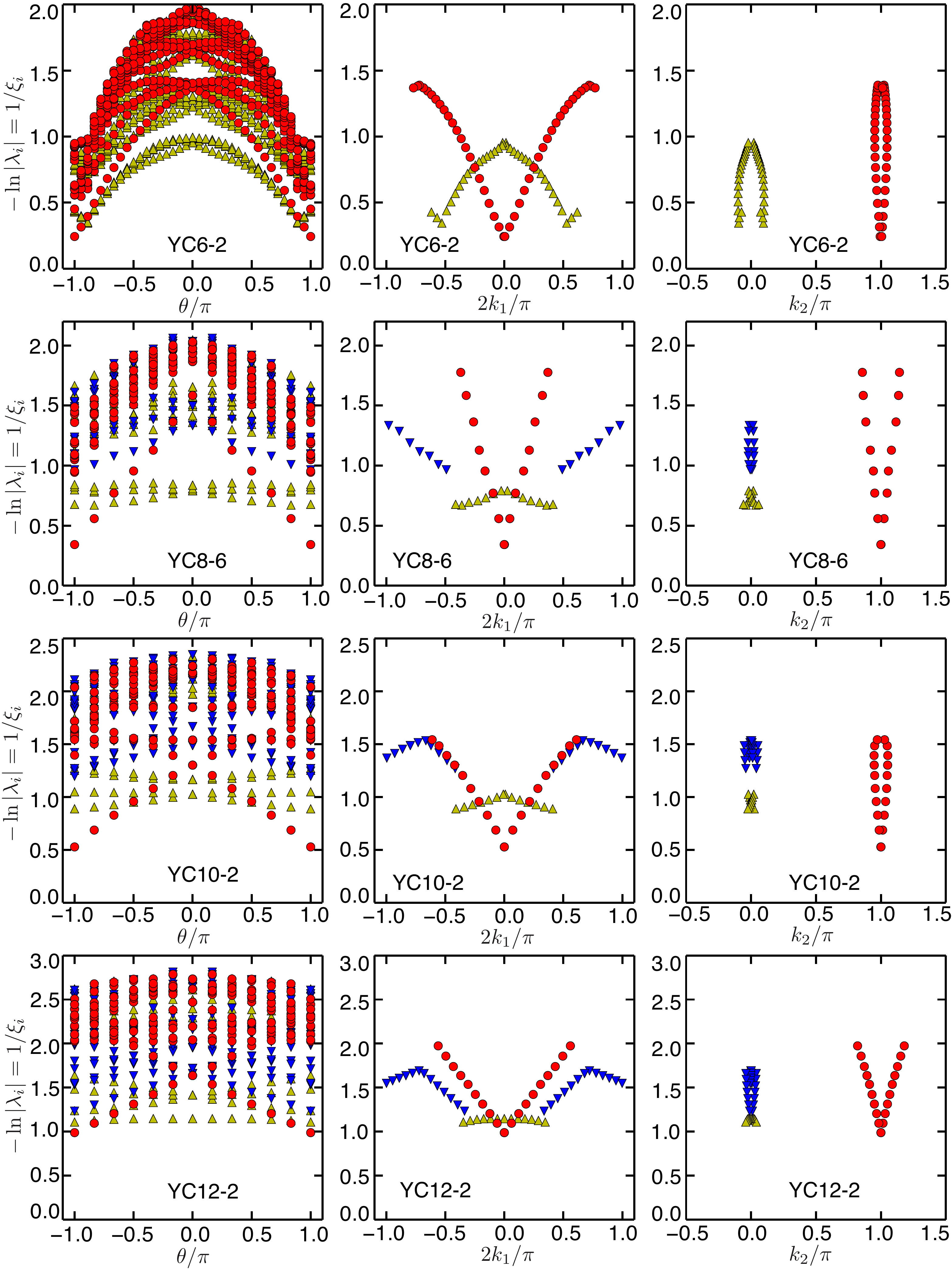}\caption{\label{fig:S1_app1} The $S^z=1$  transfer matrix  spectrum of $J_2=0.05$ for the YC6-2, YC8-6, YC10-2, YC12-2 cylinders. Here the bond dimension is $m=6000$.}
\end{figure}

\begin{figure}
\includegraphics[width=0.49\textwidth]{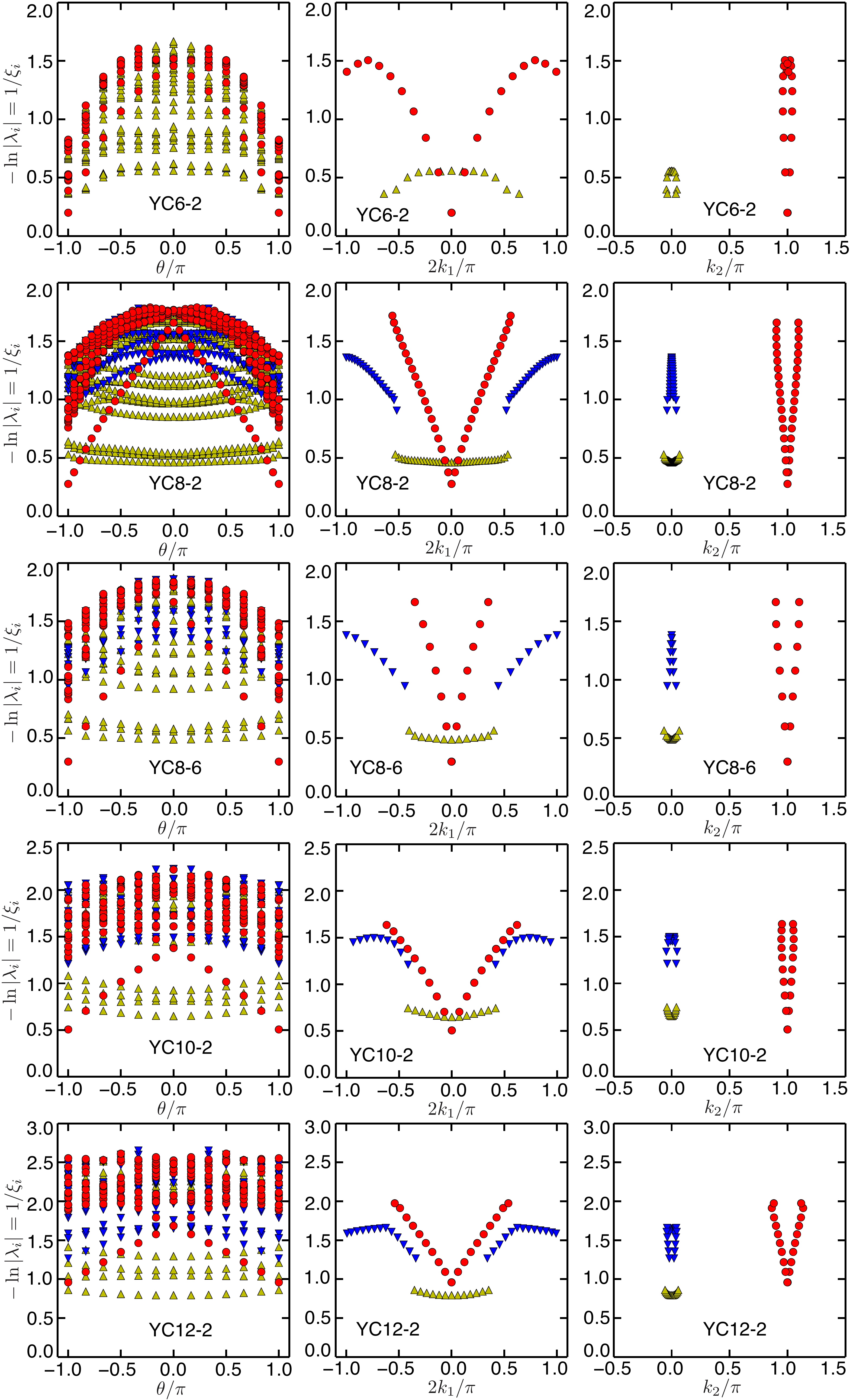}\caption{\label{fig:S1_app2} The $S^z=1$  transfer matrix  spectrum of $J_2=0.1$ for the YC6-2, YC8-2, YC8-6, YC10-2, YC12-2 cylinders. Here the bond dimension is $m=6000$.}
\end{figure}

\begin{figure}
\includegraphics[width=0.49\textwidth]{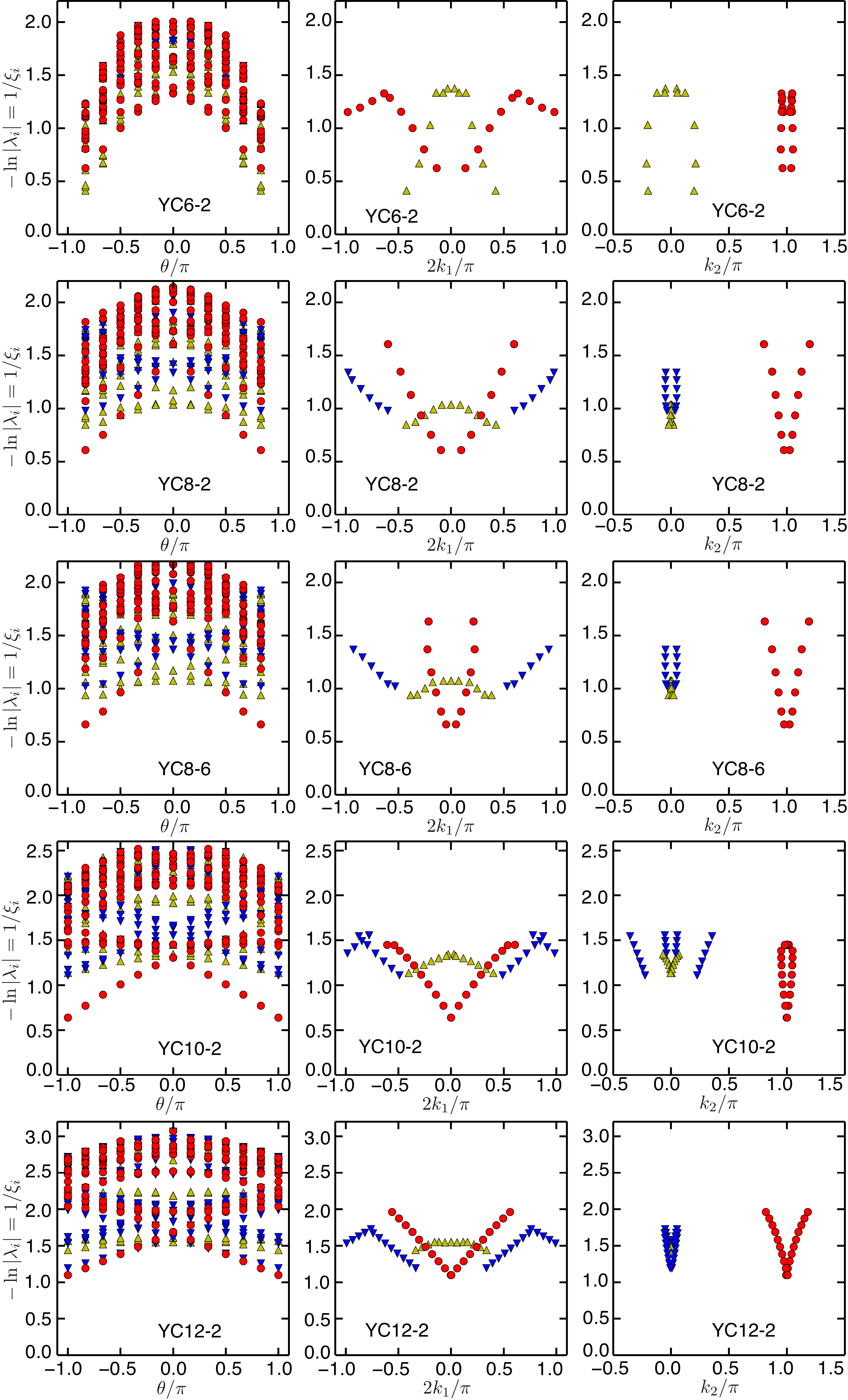}\caption{\label{fig:S1_app3} The $S^z=1$  transfer matrix  spectrum of $J_2=0$ for the YC6-2, YC8-2, YC8-6, YC10-2, YC12-2 cylinders. Here the bond dimension is $m=6000$.}
\end{figure}

\begin{figure}
\includegraphics[width=0.49\textwidth]{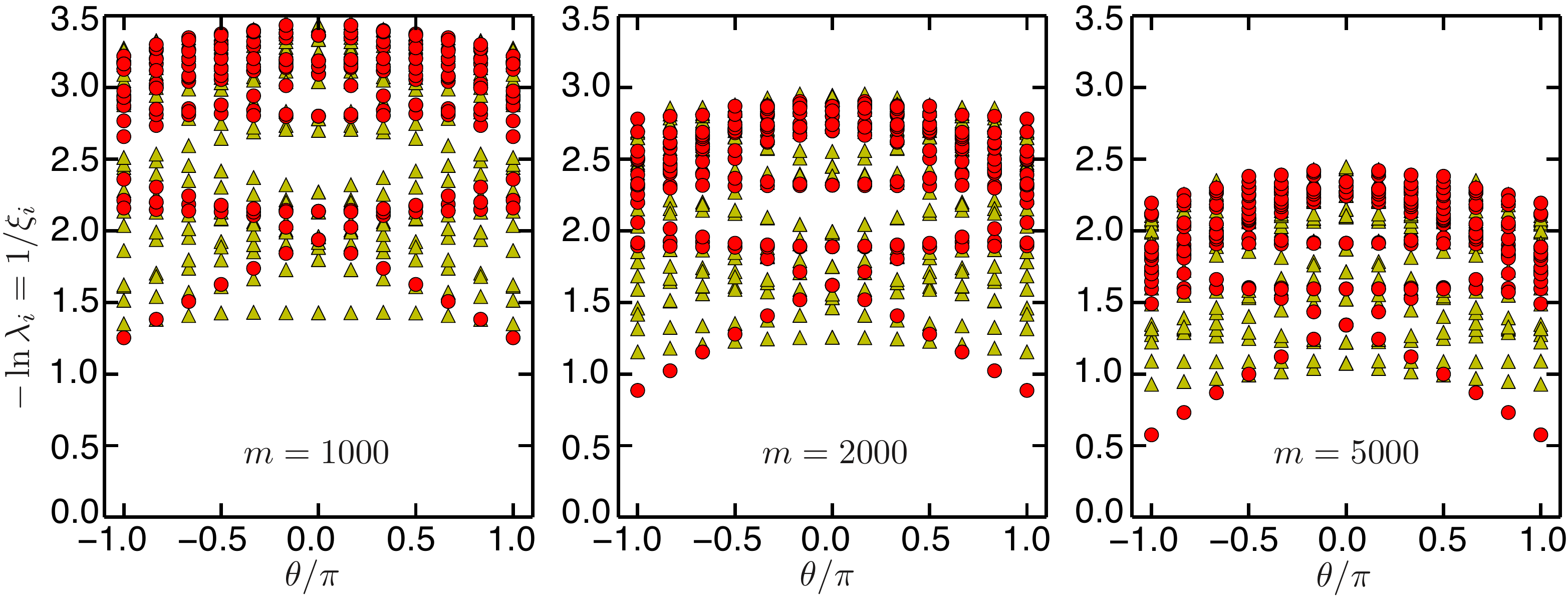}\caption{\label{fig:compare_S1_app} The dependence of the triplet spectrum on the bond dimension $m$. Here we show YC10-2, $J_2=0.05$. The red mode (red circle) is the Dirac mode, and as the bond dimension increases, it becomes much lower than the other mode (yellow triangles).}
\end{figure}

Let us first focus on the YC2n-(4k+2) cylinder.
As we showed in the main text, this class shows a clear Dirac cone structure in the $S^z=1$ sector.

We  plot the $S^z=1$ transfer matrix spectrum of $J_2=0.05$ Fig. \ref{fig:S1_app1}, $J_2=0.1$ Fig. \ref{fig:S1_app2} and $J_2=0$ Fig. \ref{fig:S1_app3} for the YC6-2, YC8-2, YC10-2, YC12-2, YC8-6 cylinders.
Clearly, the Dirac cone is independent of the $J_2$ interaction or system sizes.  
For $J_2=0$ in Fig. \ref{fig:S1_app3}, we find that it is more difficult to main the adiabaticity of the twist around $\theta=\pi$. 
Specifically, for the YC6-2 and YC8-2 cylinder, we can only adiabatically twist the system to $\theta=5\pi/6$; for $\theta=\pi$, we always end up with an ordered state with a sudden jump.
As we argued in the following section, once a Dirac spin liquid is put on a small cylinder, it might have the instability by spontaneously generating a mass gap.
Such finite size effect will be gone in the pure 2+1D limit.
This is consistent with our observation that, for a larger system size (i.e. YC10-2, YC12-2), the adiabatic twist can be maintained even for $J_2=0$.

 Comparing the excitation spectrum of different system sizes, it seems that the larger the system is, the higher the Dirac mode (red circles) is. 
 However, this is probably an artifact of the finite entanglement effect (truncation error) in our DMRG simulation. 
 Since the Dirac mode is gapless, it suffers stronger finite entanglement effect than a gapped mode.
 That is to say, the larger the truncation error is, the more we will underestimate the Dirac mode.
 This artifact can be seen in Fig. \ref{fig:compare_S1_app}, where clearly  the Dirac mode becomes lower as the bond dimension increases. 
On the other hand, to achieve same accuracy (truncation error), the required bond dimension  increases exponentially with the circumference of the cylinder.
Table \ref{Table:truncation_error} shows the truncation error for different bond dimensions, system sizes and twist angles.
For the large system size (YC10-2, YC12-2), $m=6000$ roughly gives comparable accuracy as $m=1000$ for YC8-2.
Therefore, more care should be taken if one wants to compare the results between different system sizes.

\begin{table}
\setlength{\tabcolsep}{0.12cm}
\renewcommand{\arraystretch}{1.4}
\caption{\label{Table:truncation_error}The truncation error for $J_2=0.05$. 
The truncation error increases as the system sizes.
It is also worth noting that the $\theta=\pi$ has much larger truncation error than the $\theta=0$ due to the fact that $\theta=\pi$ is likely to be gapless.}
\begin{tabular}{ccccccc}
\hline \hline
				& $m=1000$		& $m=2000$			& $m=4000$			& $m=6000$ \\ \hline
YC6-2, $\theta=0$ 	& $2.5\times10^{-7}$	&$3.1\times10^{-8}$ &$3.0\times10^{-9}$	&$5.5\times10^{-10}$\\ 
YC6-2, $\theta=\pi$  & $4.1\times10^{-6}$	&$1.3\times10^{-6}$ &$3.4\times10^{-7}$	&$1.6\times10^{-7}$\\
\hline
YC8-2, $\theta=0$	& $1.2\times10^{-5}$&$3.7\times10^{-6}$&$1.0\times10^{-6}$	&$4.6\times10^{-7}$\\
YC8-2, $\theta=\pi$	& $2.3\times10^{-5}$&$1.0\times10^{-5}$&$4.4\times10^{-6}$	&$2.8\times10^{-6}$\\
\hline
YC10-2, $\theta=0$ 	& $6.0\times10^{-5}$&$2.5\times10^{-5}$&$1.1\times10^{-5}$	&$6.2\times10^{-6}$\\
YC10-2, $\theta=\pi$  &$7.0\times10^{-5}$&$3.4\times10^{-5}$&$1.6\times10^{-5}$	& $1.1\times10^{-5}$\\
\hline
YC12-2, $\theta=0$ 	& $1.1\times10^{-4}$&$7.3\times10^{-5}$&$4.0\times10^{-5}$	&$2.7\times10^{-5}$\\
YC12-2, $\theta=\pi$  &$1.2\times10^{-4}$&$7.7\times10^{-5}$&$4.5\times10^{-5}$	&$3.1\times10^{-5}$\\
\hline \hline
\end{tabular}
\end{table}

\begin{figure}
\includegraphics[width=0.49\textwidth]{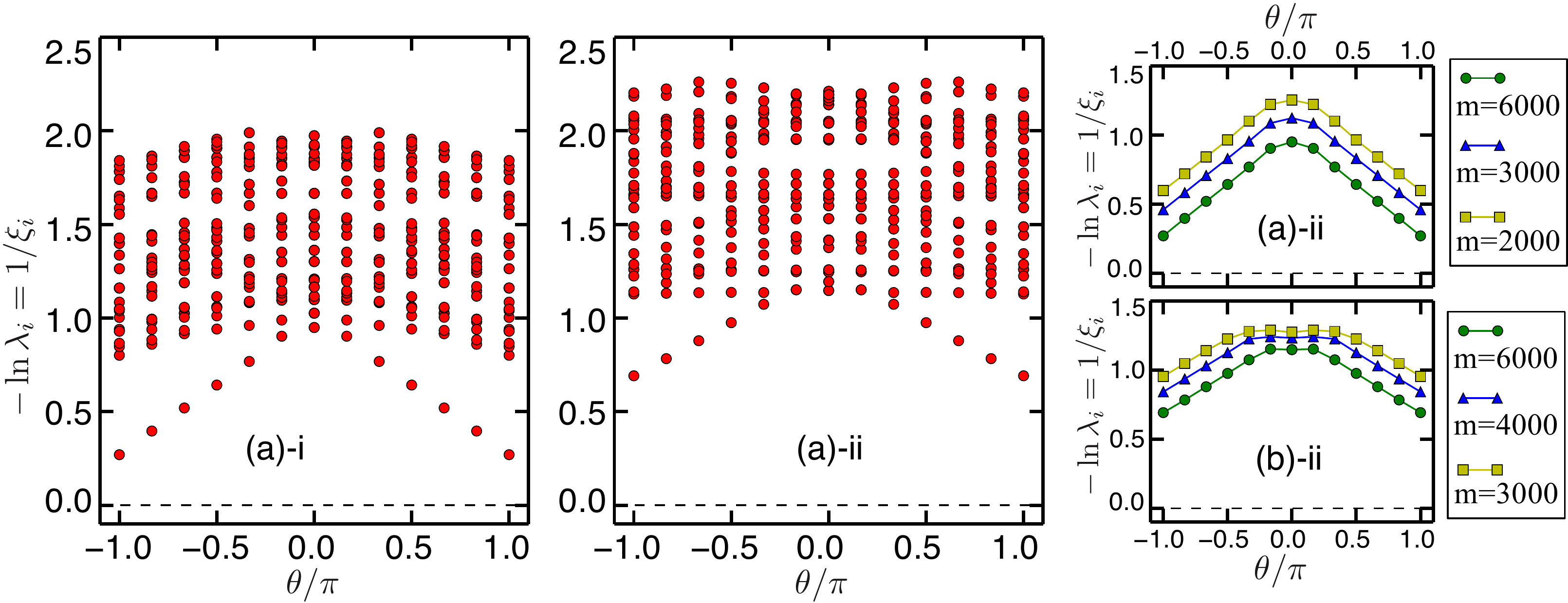}\caption{\label{fig:S0_twist_app} (a) The singlet excitation spectrum for the (a)-i YC10-2 and (a)-ii YC12-2 cylinders, here $J_2=0.05$, bond dimension $m=6000$.
(b) The dependence on the bond dimension $m$ of the lowest singlet excitation spectrum for the  (b)-i YC10-2 and (b)-ii YC12-2 cylinders.
The singlet excitation keeps on going down as the bond dimension $m$ increases.}
\end{figure}

Besides the $S^z=1$ transfer matrix spectrum, it is also interesting to look at the $S^z=0$ transfer matrix spectrum, Fig. \ref{fig:S0_twist_app}. 
As we discussed in the main text, those  $S^z=0$ excitations are more critical than the  $S^z=1$ excitations. 
For a large system size, the singlet excitation is not gapless at the Dirac point $\theta=\pi$.
We think this is again the artifact of finite bond dimension, as one can see that the singlet excitation keeps on going down as the bond dimension $m$ increases, as shown in Fig. \ref{fig:S0_twist_app}(b).

\subsubsection{YC2n-4k cylinder}

At last, let's look at the YC2n-4k cylinder. 
As we discussed in the main text, this class of cylinder behaves very different for YC2n-(4k+2) that is discussed above.
For the $\pi$-flux Dirac spin liquid, the YC2n-(4k+2) cylinder will hit the gapless Dirac point at $\theta=\pi$, while YC2n-4k cylinder will hit the gapless Dirac point at $\theta=2\pi$.
Our simulation on the YC2n-4k cylinder is also consistent with this scenario, namely the adiabaticity of the twist can be maintained after $\theta=\pi$ until $\theta\approx 4\pi/3$, after which the system collapses to the other topological sector.

\begin{figure}
\includegraphics[width=0.49\textwidth]{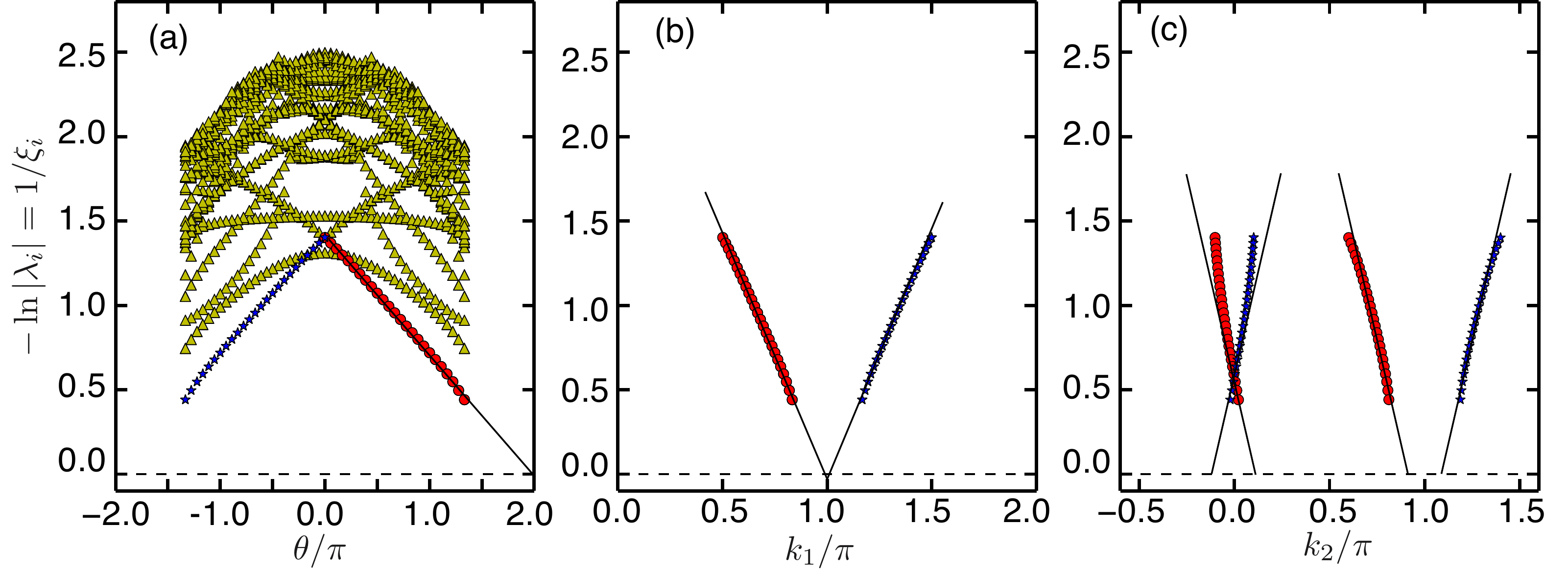}\caption{\label{fig:YC8-0,S1} The $S^z=1$ transfer matrix spectrum of YC8-0, here $J_2=0$ and the bond dimension is $m=6000$.}
\end{figure}

When the $\pi$-flux DSL hits the Dirac points at $\theta=2\pi$ on the YC2n-4k cylinder, the four Dirac fermions will be simultaneously gapless.
Those four gapless Dirac fermions could then form different gapless fermion bilinears giving rise to gapless triplet (singlet excitations).
This is again sharply distinct from YC2n-(4k+2) cylinder, where only two Dirac fermions are gapless when the system hits the Dirac points (at $\theta=\pi$).
 Therefore for the $\pi$-flux DSL, we expect that gapless triplet excitation at all the three M points will be gapless on the YC2n-4k cylinder. 
 However, we are working at a small cylinder, such that some lattice symmetry (e.g. $C_3$) is explicitly broken.
 Therefore, it is possible that the gapless triplet excitation at certain M point will be pushed to a higher energy level.

 Fig. \ref{fig:YC8-0,S1} shows the  $S^z=1$ transfer matrix spectrum of the cylinder YC8-0 with $J_2=0$.
 We find the lowest modes behaves like the Dirac modes.
 Similar as the YC2n-(4k+2) cylinder, the lowest modes show a linear dependence with the twist angle $\theta$.
 These ``Dirac modes" are actually two-fold degenerate, and they have the same $k_1$ but distinct $k_2$.
Since our simulation cannot adiabatically twist to the Dirac points, we cannot unambiguously determine the momentum of the Dirac points. 
But there are several indications that, the two Dirac modes correspond to the  $M_1=(\pi, 0)$ and $M_3=(\pi,\pi)$ points (labeled by $(k_1, k_2)$).
First,  the YC8-0 cylinder has a reflection symmetry (with the reflection axis perpendicular to $\vec a_1$), under which $M_1$ transforms to $M_3$.
Second, by doing a simple linear extrapolation for momentum $k_1$ (Fig. \ref{fig:YC8-0,S1}(b)) and $k_2$ (Fig. \ref{fig:YC8-0,S1}(c)), the momentum is consistent with $M_1$ and $M_3$.
We note that the extrapolation for $k_2$ gives $k_2\approx\pm 0.1\pi$ and $k_2\approx\pi \pm 0.1 \pi$, which has considerable discrepancy from $k_2=0$ and $k_2=\pi$.
 This discrepancy might come from the finite size effect, for example there is  scattering (momentum transfer) between two modes.
 
 \section{Instability of a Dirac spin liquid in the quasi-one dimensional limit \label{sec:instability}}
\subsection{Instability on the YC2n-(4k+2) cylinder}

\begin{figure}
\includegraphics[width=0.35\textwidth]{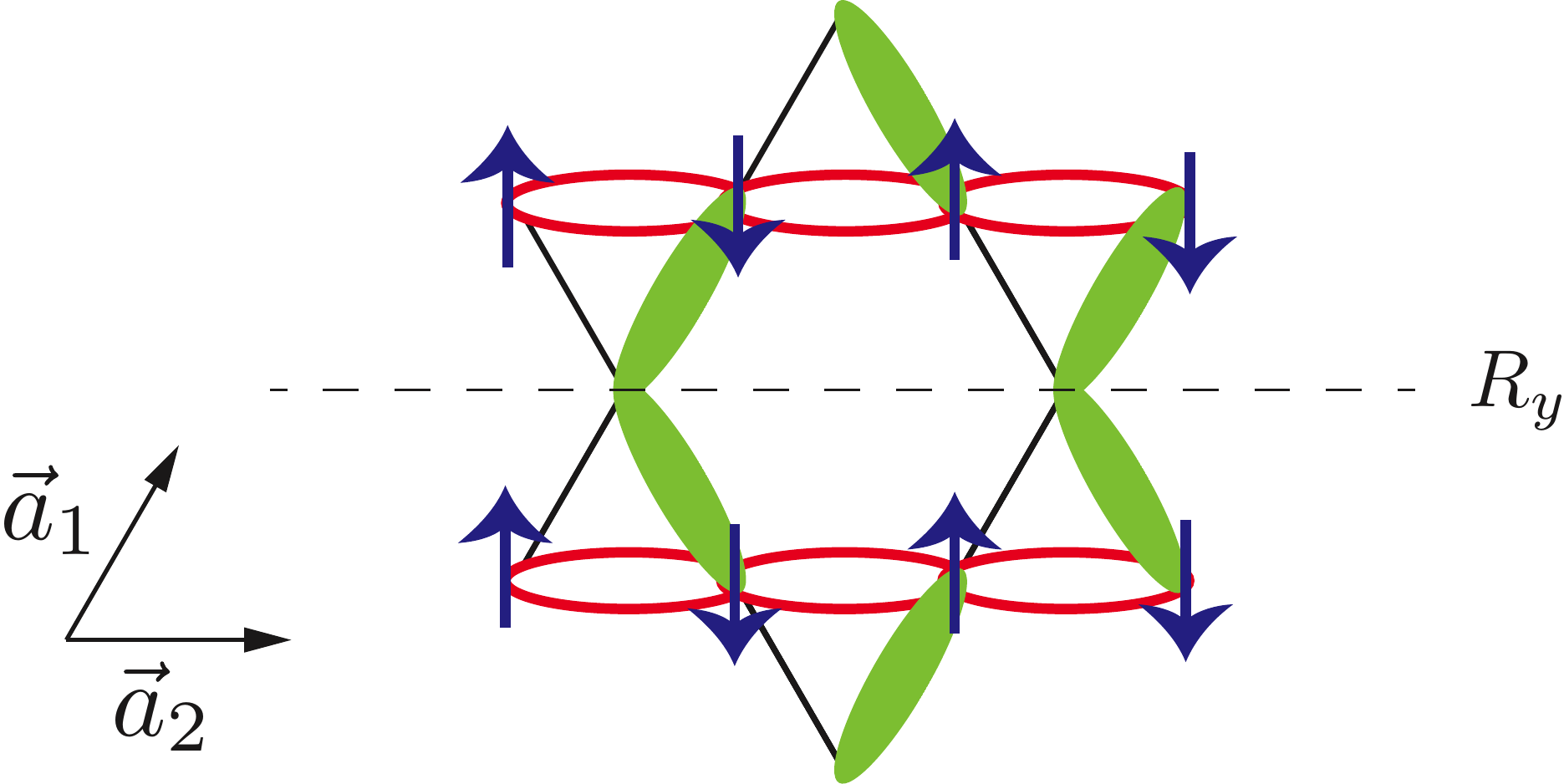}\caption{\label{fig:order_instability} The kagome spin liquid on the YC8-2 cylinder is unstable to an ordered state when $\theta=\pi$. The ordered state spontaneously breaks the spin flip symmetry $P_x=\prod (2 S^x)$ (hence the time-reversal symmetry) with forming a staggered Ising magnetization ($\langle S_i^z\rangle \approx \pm 0.03$) as represented by arrows; and breaks lattice symmetry, with bond correlations $\langle \vec S_i \cdot \vec S_j \rangle $ to be $-0.26$ (green solid bonds), $-0.22$ (red hollow bonds), and $-0.17$ (black thin bonds).}
\end{figure}

For the YC8-2 cylinder, once we tune the flux to $\theta=\pi$ in order to exactly hit the Dirac points, the kagome spin liquid is unstable to an ordered state.
Such ordered state breaks spin flip symmetry $P_x=\prod (2 S^x)$ and the lattice symmetry (e.g. $T_{a_1}$ and $C_6$), its order pattern is shown in Fig. \ref{fig:order_instability}.
Interestingly, such ordered state actually preserves certain symmetries, which are i) the reflection symmetry $R_y$, ii) translational symmetry $T_{a_2}$, iii) the combination of spin flip and translational symmetry $P_x T_{a_1}$, iv) $XY$ spin rotation symmetry.

One can work out the transformations for the Dirac fermions $\Psi$ \cite{Hermele2008},
\begin{eqnarray}
& T_{a_1}:  \Psi \rightarrow (i\mu^2) \Psi \\
 &T_{a_2}: \Psi \rightarrow (i\mu^3) \Psi \\
 &P_x:  \Psi \rightarrow (\sigma^1) \Psi \\
 &R_y: \Psi \rightarrow (\tau^{12})(i \mu^{12}) \Psi
 \end{eqnarray}
where $ \tau^{12}=(\cos \frac{\pi}{6} \tau^1-\sin \frac{\pi}{6} \tau^2)$ and $\mu^{12}= (\cos \frac{\pi}{4} \mu^1-\sin \frac{\pi}{4}\mu^2)$, $\bar \Psi = i \Psi^\dag \tau^3$. 
$\tau$ represents the spinor index of Dirac fermions, $\sigma$ represents the spin index, $\mu$ represents the valley index.
With the above symmetry transformation rules, one can straightforwardly find out  the mass term (of the Dirac fermions) that gives  the order pattern (Fig. \ref{fig:order_instability}) is $i \bar \Psi \sigma^3 \mu^3 \Psi$.

\subsection{Spontaneous mass generation of Dirac fermions in 1+1 dimension}

As we discussed in Sec.~\ref{sec:KAH_results},
when the kagome spin liquid is tuned to exactly hit the Dirac points  on a small cylinder (YC2n-(4k+2) geometry), we numerically find an
instability toward an ordered state by spontaneously generating a mass term  $i \bar \Psi \sigma^3 \mu^3 \Psi$.
This immediately raises the question that, whether this results imply that the U(1) DSL ($N_f=4$ QED3) is unstable to the spontaneously chiral symmetry breaking (CSB).
The CSB of QED3 is still an open issue, and it is unclear whether $N_f=4$ QED3 will eventually flow to an interacting conformal fixed point or not \cite{Grover2014}.
Our numerical simulation, on the other hand, was carried on a quasi-1D cylindrical geometry, for which the issue of spontaneous mass generation is different from the 2+1D limit.
The differences are two-fold: i) The U(1) gauge field is more gentle in quasi-1D, it simply reduces the $N_f$ flavors of 1D Dirac fermions to $N_f-1$ coupled Tomonaga-Luttinger liquids (TLL)~ \cite{Sheng2009}.
ii) The effect of four-fermion interactions on Dirac fermions are more drastic in 1D than that in 2D, namely in 2D all four-fermion interactions are irrelevant for (free) Dirac fermions; while in 1D, four-fermion interactions might be relevant or marginally relevant.

For the U(1) DSL on the YC2n-(4k+2) cylinder with $\theta=\pi$-flux, two Dirac fermions are gapless. Then the dynamical $U(1)$ gauge field reduces the system to a TLL with central charge $c=1$, described by
\begin{equation}
H=\frac{v}{2} \left( \frac{1}{K} (\partial_t \phi)^2+ K (\partial_x \phi)^2\right).
\end{equation}
$K$ is the Luttinger parameter.
Generically, there are perturbations to  that can potentially gap out the system. 
Assuming that the U(1) symmetry is preserved,
the most relevant perturbation is $\lambda \cos 2\phi$, which is irrelevant when $K>1/2$~\cite{Giamarchi_1Dbook}.
For $K<1/2$, the perturbation is relevant, and will gap out the Luttinger liquid, which corresponds to the mass generation of the Dirac fermions.
For $K=1/2$, the perturbation is marginally relevant or marginally irrelevant, depending on the sign of $\lambda$.
Microscopically, it is difficult to extract the Luttinger parameter $K$. 
However, for the $c=1$ Luttinger liquid with $SU(2)$ symmetry, the $K$ is fixed to $K=1/2$.
Our system is very close to this limit (the twist boundary condition slightly breaks $SU(2)$ symmetry), hence we expect $K\approx 1/2$.
Consequently, it is reasonable that in our numerical simulation on a small YC2n-(4k+2) (e.g. YC8-2) cylinder, the DSL is unstable to an ordered state with a mass term spontaneously generated.
\bibliography{spin_liquid.bib}
\end{document}